\documentclass[10pt]{article}

\usepackage{bm}
\usepackage{fullpage}  
\usepackage{url}
\usepackage{algorithm, algorithmicx,algpseudocode}
\usepackage{multirow} 
\usepackage{hhline}  

\usepackage{verbatim}

\usepackage{graphicx}
\usepackage{amsmath,amssymb,amsfonts,graphicx,amsthm,mathtools,nicefrac}
\usepackage{lscape}
\usepackage{color}
\usepackage{authblk}
\usepackage{subfigure}
\usepackage{rotating}

\allowdisplaybreaks
\newcommand\numberthis{\addtocounter{equation}{1}\tag{\theequation}}

\newtheorem{theorem}{Theorem}[section]
\newtheorem{assumption}{Assumption}[section]
\newtheorem{remark}{Remark}[section]

\numberwithin{equation}{section}

\newtheorem{lemma}[theorem]{Lemma}

\DeclareMathOperator{\rank}{\mathrm{rank}}
\DeclareMathOperator{\diag}{\mathrm{diag}}
\DeclareMathOperator{\dist}{\mathrm{dist}}

\DeclareMathOperator{\trace}{trace}

\def\la{\left\langle}
\def\ra{\right\rangle}
\def\lb{\left(}
\def\rb{\right)}
\def\lcb{\left\{}
\def\rcb{\right\}}

\def\lsb{\left[}
\def\rsb{\right]}
\def\lab{\left|}
\def\rab{\right|}

\newcommand{\vecnorm}[2]{\left\| #1\right\|_{#2}}

\newcommand{\matsnorm}[2]{\left\| #1\right\|_{{#2}}}

\newcommand{\fronorm}[1]{\ensuremath{\matsnorm{#1}{\footnotesize{\mathsf{F}}}}}
\newcommand{\opnorm}[1]{\ensuremath{\matsnorm{#1}{}}}
\newcommand{\twoinf}[1]{\ensuremath{\matsnorm{#1}{\footnotesize{\mbox{2,$\infty$}}}}}

\newcommand{\twonorm}[1]{\ensuremath{\matsnorm{#1}{\footnotesize{\mbox{2}}}}}

\newcommand{\bfm}[1]{\bm{#1}}

\newcommand{\E}[2][]{\mathbb{E}_{#1} \left[ #2 \rule{0mm}{3mm}\right]}

\def\va{\bfm a}   \def\mA{\bfm A}  
\def\vb{\bfm b}   \def\mB{\bfm B}  
     \def\C{\mathbb{C}}
   \def\mD{\bfm D}  
\def\ve{\bfm e}     
    
\def\vg{\bfm g}   \def\mG{\bfm G}  
\def\vh{\bfm h}     
   \def\mI{\bfm I}  
   \def\mJ{\bfm J}  
   \def\mK{\bfm K}  
   \def\mL{\bfm L}  
   \def\mM{\bfm M}

   \def\mP{\bfm P}  
   \def\mQ{\bfm Q}  
   \def\mR{\bfm R}  \def\R{\mathbb{R}}
   \def\mS{\bfm S}  
     
   \def\mU{\bfm U}  
   \def\mV{\bfm V}  
\def\vw{\bfm w}     
\def\vx{\bfm x}   \def\mX{\bfm X}  
\def\vy{\bfm y}   \def\mY{\bfm Y}  
\def\vz{\bfm z}   \def\mZ{\bfm Z}

\def\calA{{\cal  A}}

\def\calD{{\cal  D}}

\def\calG{{\cal  G}} 
\def\calH{{\cal  H}} 
\def\calI{{\cal  I}}

\def\calM{{\cal  M}} 
 
\def\calO{{\cal  O}} 
\def\calP{{\cal  P}}

\def\bzero{\bfm 0}

\newcommand{\bfsym}[1]{\bm{#1}}

           \def\bDelta {\bfsym {\Delta}}

                 \def\bPhi {\bfsym {\Phi}}

             \def\bSigma{\bfsym \Sigma}

            \def\bPsi {\bfsym {\Psi}}

\def\bPsi {\bfsym {\Psi}}

              \def\hbSigma{\widehat{\bfsym \Sigma}}

\def \hM{\widehat{\mM}}
\def \hU{\widehat{\mU}}
\def \hV{\widehat{\mV}}

\def \hZ{\widehat{\mZ}}

\def \hL{\widehat{\mL}}
\def \hR{\widehat{\mR}}

\def \calGT{\calG^{\ast}}
\def \GGT{\calG\calGT}
\def \calAT{\calA^{\ast}}

\def \mLQ {\mL^\natural \mQ }
\def \mRQ {\mR^\natural \mQ }

\def \tran {\mathsf{T}}
\def \tranH{\mathsf{H}}

\def \bzero{\bm 0}

\begin{document}
\title{Blind  Super-Resolution  of  Point Sources  via  Projected  Gradient  Descent}
\author{Sihan Mao and Jinchi Chen
\thanks{This work was supported by National Science Foundation of China under Grant No. 12001108. {\color{black}Parts of the results in this paper will be presented at the 2022 IEEE International Symposium on Information Theory \cite{mao2021blind}.	}
 
}
\thanks{
The authors are with School of Data Science, Fudan University, Shanghai, China (email: 18110980008@fudan.edu.cn; jcchen.phys@gmail.com)
}}


\maketitle

\begin{abstract}
Blind super-resolution can be cast as a low rank matrix recovery problem by exploiting the inherent simplicity of the signal and the low dimensional structure of point spread functions. In this paper, we develop a simple yet efficient non-convex projected gradient descent method for this problem based on the low rank structure of the vectorized Hankel matrix associated with the target matrix. Theoretical
analysis indicates that the proposed method exactly converges to the target matrix with a linear convergence rate under the similar conditions as convex approaches. Numerical  results show that our approach is competitive with existing convex approaches in terms of recovery ability and efficiency.
\end{abstract}


 \section{Introduction}
Blind super-resolution is the problem of estimating high-resolution information of a signal from its low-resolution measurements when the point spread functions (PSFs) are unknown. Such problem arises in a wide variety of applications, including seismic data analysis \cite{margrave2011gabor}, nuclear magnetic resonance spectroscopy \cite{qu2015accelerated}, multi-user communication system \cite{luo2006low}, and 3D single-molecule microscopy \cite{quirin2012optimal}. In particular, when the knowledge of PSFs is available, blind super-resolution reduces to the super-resolution problem \cite{candes2013super, candes2014towards}. 

 Without any additional assumptions, blind super-resolution of point sources is an ill-posed problem. To alleviate this issue, it is common to assume that the PSFs belong to a known low-dimensional subspace. Under this assumption and utilizing the lift technique, blind super-resolution of point sources can be formulated as a matrix recovery problem. By exploiting low dimensional structures of the target matrix, a series of works \cite{chi2016guaranteed, yang2016super, li2019atomic, chen2020vectorized, suliman2021mathematical}  theoretically studied under which conditions the target matrix can be recovered. The author in \cite{chi2016guaranteed} considered the setting where the PSF is shared among all point sources, and established the recovery guarantees for the atomic norm minimization {(ANM)} method. Yang et al. \cite{yang2016super} further studied the same method, but with multiple unknown PSFs. Li et al. \cite{li2019atomic} provided robust analysis of blind 1D super-resolution and later the work in \cite{suliman2021mathematical} generalized  \cite{li2019atomic}  to 2D case. {\color{black}Recently, Chen et al. \cite{chen2020vectorized} proposed a nuclear norm minimization method based on the vectorized Hankel lift framework, which also appears in \cite{yang2018sparse,zhang2018multichannel} but for matrix completion.} Moreover, \cite{chen2020vectorized} established the corresponding exact recovery guarantees. While strong theoretical guarantees have been built for blind super-resolution based on convex methods, these approaches are computational inefficient for the high dimensional setting. Therefore it is necessary to design efficient and provable algorithms to deal with the large-scale regime.
 
{\color{black}
In the past few years, substantial progress has been made on designing and analyzing provable fast algorithms for applications from science and engineering via non-convex optimization \cite{jain2017non, chen2018harnessing, chi2019nonconvex}, including matrix completion \cite{chen2015fast, zheng2016convergence},  phase retrieval \cite{shechtman2015phase}, blind deconvolution \cite{li2019rapid},  spectrally sparse signal recovery \cite{cai2018spectral, cai2019fast}, to name just a few. The goal of this work is to develop an efficient non-convex algorithm for blind super-resolution problem.

 \subsection{Comparisons with Related Work and Main Contributions}

 Our work is closely related to \cite{chen2020vectorized, cai2018spectral, zhu2021low}. 
 As already mentioned,
 \cite{chen2020vectorized} proposed a convex approach called Vectorized Hankel Lift (VHL) for blind super-resolution. 
 Based on this framework, 
 we develop an efficient and provable non-convex algorithm for blind super-resolution of point sources.
 More precisely, we parameter the vectorized Hankel matrix corresponding to a candidate  solution  in terms of the Burer--Monteiro factorization and develop a projected gradient descent method  to directly recover the low-rank factors. 
 
 Our algorithm is inspired by the method in \cite{cai2018spectral}, where the projected gradient descent was developed for spectrally sparse signal recovery problem based on the low rank structure of the Hankel matrix corresponding to the target signal. Despite this,
 both the structures of sensing operator and target matrix in this paper are substantially different from that in \cite{cai2018spectral}. Therefore, the convergence analysis in \cite{cai2018spectral} can not be easily extended to our model. 
 
 Recently, \cite{zhu2021low} follows our work and develops an iterative hard thresholding method based on the framework of vectorzied Hankel lift to solve blind super-resolution problem. It is worth pointing out that
 they directly apply our result to bound the initialization error of their method. 
 Furthermore, their proof idea is inspired by the guarantee analysis of low rank matrix recovery over Riemannian manifold \cite{cai2019fast, wei2016guarantees}. Therefore the proof techniques are totally different with ours.

  }
 
 
 {\color{black} The main contributions of this work are summerized as follows. Firstly, we present a new non-convex algorithm called projected gradient descent via vectorized Hankel lift (PGD--VHL) for blind super-resolution. Numerical experiments show that PGD--VHL is competitive with  convex recovery methods such as ANM and VHL 
 in terms of recovery ability, but is much more efficient. Secondly, we establish the recovery performance of PGD--VHL. Our results show that PGD--VHL started from a spectral initialization converges linearly to the target matrix under the similar sample complexity as convex approaches. Lastly, it is worth mentioning that the theoretical guarantee of PGD--VHL requires a slightly milder assumption on the low-dimensional subspace than that for VHL in \cite{chen2020vectorized}.
 }
 
 


 
 \subsection{Organization and Notation}
 The rest of this paper is organized as follows. Section \ref{sec: problem formulation} gives the problem setup of blind super-resolution of point sources. Section \ref{subsec: pgd} presents the PGD--VHL algorithm whose the exact recovery guarantee is provided in Section \ref{sec: main result}. Numerical evaluations are presented to illustrate the performance of PGD--VHL in Section \ref{sec: numerical}. All proofs are deferred to Section \ref{sec: proof}. Finally, we conclude this paper and propose some future work in Section \ref{sec: conclusion}.
 
 Some notations used throughout this paper are presented as follows. Symbols for vectors, matrices and operators are in bold lowercase letters, bold uppercase letters and calligraphic letters, respectively. In this paper, vectors and matrices are indexed starting with zero. For a complex number $x$, its real part is denoted by $\Re (x)$. The transpose, complex
 conjugate, complex transpose, spectral norm and Frobenius
 norm of matrix $\mX$ are denoted as $\mX^\tran$, $\overline{\mX}$, $\mX^\tranH$, $\opnorm{\mX}$ and $\fronorm{\mX}$, respectively. The inner product of two matrices $\mX_1$ and $\mX_2$ is defined as $\la \mX_1, \mX_2 \ra = \trace(\mX_1^\tranH\mX_2)$. {\color{black}Moreover, we will refer to $\mA \odot \mB$ and $\mA \otimes \mB$ as the Hadamard product and Kroncker product, respectively.} We use $\vx[\ell]$ to denote the $\ell$-th entry of $\vx$ and $\mX(j,:)$ to denote the $j$th row of $\mX$. 
 Moreover, we use the MATLAB notation $\mX(i:j,k)$ to denote a vector of length $j-i+1$, with entries $\mX_{i,k},\cdots,\mX_{j,k}$.  
 The identity operator are denoted as $\calI$. Let $\calH$ be the vectorized Hankel lift operator 
 which maps a matrix $\mX\in\C^{s\times n}$ into an $sn_1\times n_2$ matrix, 
 \begin{align}
 	\label{vhl}
 	\calH(\mX) = \begin{bmatrix}
 		\vx_0 & \vx_1 &\cdots & \vx_{n_2-1}\\
 		\vx_1 & \vx_2 &\cdots & \vx_{n_2}\\
 		\vdots& \vdots &\ddots &\vdots\\
 		\vx_{n_1-1}&\vx_{n_1}&\cdots &\vx_{n-1}\\
 	\end{bmatrix}\in\C^{sn_1\times n_2},
 \end{align}
 where $\vx_i\in\C^s$ is the $i$-th column of $\mX$ and $n_1+n_2 = n+1$.  We denote the adjoint of $\calH$ by $\calH^\ast$, which is a linear mapping from $\C^{sn_1\times n_2}$ to $\C^{s\times n}$. In particular, for any matrix $\mZ\in\C^{sn_1\times n_2}$, the $i$-th column of $\calH^\ast(\mZ)$ is given by 
 \begin{align*}
 	\calH^\ast(\mZ)\ve_i = \sum_{\substack{ j+k = i\\ 0\leq j \leq n_1 -1,0\leq k\leq n_2-1 } }\vz_{j,k} ,
 \end{align*}
 where $\vz_{j,k}$ is the $(j,k)$-th block of $\mZ$ such that $\vz_{j,k} = \mZ(js : (j + 1)s - 1, k)$. Letting $\calD^2 = \calH^\ast \calH$, we have
 \begin{align*}
 	\calD^2(\mX) = \begin{bmatrix}
 		w_0\vx_0 & \cdots & w_{n-1} \vx_{n-1}
 	\end{bmatrix},
 \end{align*}
 where the scale $w_i$ is defined as 
 \begin{align*}
 	w_i = \#\{(j,k) | j+k=i, 0\leq j \leq n_1-1, 0\leq k\leq n_2-1 \}.
 \end{align*}
 Moreover, we define  $\calG = \calH\calD^{-1}$. The adjoint of $\calG$ denoted $\calGT$ is given by $\calGT = \calD^{-1}\calH^\ast$. Additionally, $\calG$ and $\calG^\ast$ satisfy
 \begin{align}
 	\label{property of G}
 	\calGT\calG = \calI,\quad\quad
 	\opnorm{\calG}=1,\quad\mbox{and } \opnorm{\calGT}\leq 1.
 \end{align}
 We use $\mG_i$ to denote the matrix defined by
 \begin{align}
 	\label{eq: Hankel basis}
 	\mG_i = \frac{1}{\sqrt{w_i}}\sum_{\substack{j+k=i\\0\leq j\leq n_1-1, 0\leq k \leq n_2-1}} \ve_j\ve_k^\tran.
 \end{align}
 Then one has
 \begin{align}
 	\label{eq: calG}
 	\calG(\mX)=\sum_{i=0}^{n-1}\calG \lb\vx_i\ve_i^\tran\rb = \sum_{i=0}^{n-1}\mG_i\otimes\vx_i,
 \end{align}
 where $\mG_i\otimes\vx_i$ denotes the Kronecker product between $\mG_i$ and $\vx_i$.
 
 Throughout this paper, $c, c_0,c_1,\cdots$ denote absolute positive numerical constants whose values may vary from line to line. The notation $n=\calO(m)$ means that there exists an absolute constant $c>0$ such that $n\leq cm$.

 \section{Problem formulation}
 \label{sec: problem formulation}
 The point source signal model can be represented as a superposition of $r$ spikes
 \begin{align}
 	\label{eq: model form}
 	x(t) = \sum_{k=1}^r d_k\delta(t-\tau_k),
 \end{align}
 where $\delta(\cdot)$ is the Dirac function, {\color{black}{$d_k\in\C$ and $\tau_k\in[0,1)$}} are the amplitude and location of the $k$-th point source, respectively. Let $\{g_k(t)\}_{k=1}^r$ be the unknown point spread functions depending on the locations of point sources. The observation is a convolution between $x(t)$ and $\{g_k(t)\}_{k=1}^r$, that is, 
 \begin{align}
 	\label{eq: sample in time domain}
 	y(t) = \sum_{k=1}^r d_k\delta(t-\tau_k) * g_k(t) = \sum_{k=1}^r d_k\cdot g_k(t-\tau_k).
 \end{align} 
 After taking the Fourier transform and sampling, we obtain the measurements as 
 \begin{align}
 	\label{observation model}
 	\vy[j] = \sum_{k=1}^r d_k e^{-2\pi \imath \tau_k\cdot j} \hat{g}_k[j],\quad j=0,\cdots n-1.
 \end{align}
 Let $\vg_k = \begin{bmatrix} \hat{g}_k[0] &\cdots & \hat{g}_k[n-1]\end{bmatrix}^\tran$ be a vector corresponding to the $k$-th unknown point spread function. The goal is to estimate $\{d_k, \tau_k\}_{k=1}^r$ as well as $ \{\vg_k\}_{k=1}^r$ from \eqref{observation model}. 
 
 Obviously, the problem of blind super-resolution is ill-posed without any additional assumptions, because the number of unknowns in \eqref{observation model} is $\calO(nr)$, which is larger than the number of samples $n$. To tackle this issue, we follow the same route as that in \cite{ahmed2013blind, chi2016guaranteed, yang2016super,chen2020vectorized} and assume that all the Fourier samples of the unknown PSFs $\{\vg_k\}_{k=1}^r$ belong to a known low-dimensional subspace spanned by the columns of $\mB \in \C^{n\times s}$ with $s<n$, i.e.,
 \begin{align}
 	\label{eq: low-dim of g}
 	\vg_k = \mB\vh_k,
 \end{align}
 where $\vh_k\in\C^{s}$ denotes the unknown directional vector of $\vg_k$ in the subspace.  According to the subspace assumption \eqref{eq: low-dim of g} and using the lift trick \cite{ahmed2013blind}, we can easily rewrite (\ref{observation model}) as a set of linear measurements with respect to the target matrix $\mX^\natural = \sum_{k=1}^r d_k\vh_k\va_{\tau_k}^\tran$,
 \begin{align}
 	\label{eq: lifting}
 	\vy[j] = \la\vb_j\ve_j^\tran, \mX^\natural \ra,\quad j=0,\cdots,n-1,
 \end{align}
 where $\va_{\tau_k} = \lsb 1,e^{-2\pi \iota \tau_k},\cdots,e^{-2\pi \iota \tau_k\cdot(n-1)}\rsb^\tran$, $\vb_j\in\C^{s}$ is the $j$th column vector of $\mB^\tranH$, $\ve_j$ is the $j$-th standard basis of $\R^n$. 
 The measurement model \eqref{eq: lifting} can be rewritten succinctly as
 \begin{align}
 	\label{eq: compact form}
 	\vy = \calA(\mX^\natural),
 \end{align}
 where $\calA:\C^{s\times n}\rightarrow \C^n$ is the linear operator. Let $\calAT$ be the adjoint operator of $\calA$ which is given by $\calAT(\vy) = \sum_{j=0}^{n-1}\vy[j]\vb_j\ve_j^\tran$. Furthermore, define $\mD = \diag\left(\sqrt{w_0} ,\cdots , \sqrt{w_{n-1}} \right)$. We have $\mD\calA(\mX) = \calA\calD(\mX)$ for any $\mX$. The measurements can be reformulated as 
 \begin{align}
 	\label{weighted measurements}
 	\mD\vy = \calA\calD(\mX^\natural).
 \end{align}

 Note that once the data matrix $\mX^\natural$ is reconstructed, the frequencies $\{ \tau_k\}_{k=1}^r$ can be retrieved through spatial smoothing MUSIC \cite{evans1981high,evans1982application,yang2019source,chen2020vectorized}, and the amplitudes $\{d_k\}_{k=1}^r $ and coefficients $\{\vh_k\}_{k=1}^r$ can be estimated by solving an over-determined linear system. Therefore in this work we focus on the problem of recovering $\mX^\natural$ from its linear measurements \eqref{eq: compact form}. 
 
 It has been shown that $\calH(\mX^\natural)$ is a rank-$r$ matrix \cite{chen2020vectorized} and thus the matrix $\calH(\mX^\natural)$ admits low rank structure when $r \ll \min(sn_1, n_2)$.  Equipped with the low rank structure of $\calH(\mX^\natural)$, it is natural to recover $\mX^\natural$ by solving the constrained least squares problem
 \begin{align}
 	\label{original obj}
 	\min_{\mX}~\frac{1}{2}\twonorm{\mD\vy - \calA\calD(\mX)}^2 ~\text{ s.t. } \rank(\calH(\mX)) = r.
 \end{align}
 Letting $\mZ = \calH(\mX) = \calG\calD(\mX)$ for any $\mX$, it can be verified that $(\calI - \GGT)(\mZ) = \bzero$. To eliminate the rank constrain in \eqref{original obj}, we apply the Burer--Monteiro factorization \cite{burer2003nonlinear} to parameterize $\mZ$ as $\mZ = \mL\mR^\tranH$, where $\mL\in \C^{sn_1\times r}$ and $\mR\in \C^{n_2\times r}$ are two rank-$r$ matrices. Therefore, the optimization problem \eqref{original obj} can be rewritten as 
 \begin{align}
 	\label{original obj 2}
 	\min_{{\color{black}{\mL,\mR}}}~\frac{1}{2}\twonorm{\mD\vy - \calA\calGT(\mL\mR^\tranH)}^2 \text{ s.t. } (\calI - \GGT)(\mL\mR^\tranH) = \bzero.
 \end{align}
 Before introducing our algorithm, we make an assumption that $\mZ^\natural = \calH(\mX^\natural)$ is $\mu_1$-incoherent which is defined below. 
 \begin{assumption}
 	\label{assumption 1}
 	Let $\mZ^\natural= \mU\bSigma\mV^\tranH$ be the singular value decomposition of $\mZ^\natural$, where $\mU\in\C^{sn_1\times r}, \bSigma\in\R^{r\times r}$ and $\mV\in\C^{n_2\times r}$. Denote $\mU^\tranH = \begin{bmatrix}
 		\mU_0^\tranH &\cdots &\mU_{n_1-1}^\tranH
 	\end{bmatrix}^\tranH$, where $\mU_j= \mU[j s : (j+1) s - 1, :]$ is the $j$-th block of $\mU$ for $j=0,\cdots n_1-1$. The matrix $\mZ^\natural$ is $\mu_1$-incoherent if $\mU$ and $\mV$ obey that
 	\begin{align*}
 		\max_{0\leq j\leq n_1-1} \fronorm{\mU_j}^2 \leq \frac{\mu_1 r}{n} \text{ and }\max_{0\leq k \leq n_2-1} \vecnorm{\ve_k^\tran\mV}{2}^2 \leq \frac{\mu_1 r}{n}
 	\end{align*}
 	for some positive constant $\mu_1 $.
 \end{assumption}
 \begin{remark}
 	Assumption \ref{assumption 1} is the same as the one made in \cite{candes2009exact,zhang2018multichannel} for low rank matrix recovery and is used in \cite{chen2020vectorized} for blind super-resolution. It has been established that Assumption \ref{assumption 1} is obeyed when the minimum wrap-up distance between the locations of point sources is greater than about $2/n$. 
 \end{remark}
 Let $\mu$ and $\sigma$ be two numerical constants such $\mu_1 \leq \mu$ and $\sigma_1\leq \sigma$, and $\calM$ be a convex set defined as follows
 \begin{align}
 	\label{eq: convex set}
 	\calM = \bigg\{ \begin{bmatrix}
 		\mL\\
 		\mR\\
 	\end{bmatrix}~:~ \max_{0\leq j\leq n_1-1} \fronorm{\mL_j} \leq \sqrt{\frac{\mu r\sigma }{n}},  \twoinf{\mR} \leq  \sqrt{\frac{\mu r\sigma}{n} }\bigg\}, 
 \end{align}
where $\mL_j$ is the $j$-th block of $\mL$. Define 
 \begin{align*}
 	\mM^\natural = \begin{bmatrix}
 		\mL^\natural \\
 		\mR^\natural\\
 	\end{bmatrix} = \begin{bmatrix}
 		\mU\bSigma^{1/2}\\
 		\mV\bSigma^{1/2}
 	\end{bmatrix}.
 \end{align*}
 Since $\mZ^\natural$ is $\mu_1$-incoherent, we have $\mM^\natural\in\calM$. Therefore, we consider a penalized version of \eqref{original obj 2} for recovering the factorized matrices:
\begin{align}
\label{obj function}
\min_{\mM\in\calM }~\bigg\{ f(\mM):=\frac{1}{2} \vecnorm{\mD\vy - \calA\calGT(\mL\mR^\tranH)}{2}^2 + \frac{1}{2}\fronorm{\left( \calI - \GGT \right)(\mL\mR^\tranH)}^2+ \frac{1}{16}\fronorm{\mL^\tranH\mL - \mR^\tranH\mR}^2\bigg\},
\end{align}
 where $\mM = \begin{bmatrix}
 	\mL^\tranH &\mR^\tranH
 \end{bmatrix}^\tranH\in\C^{(sn_1+n_2)\times r}$, and the last term penalizes the mismatch between $\mL$ and $\mR$, which is widely used in rectangular low rank matrix recovery \cite{tu2016low, zheng2016convergence,chi2019nonconvex}.  
 \section{Algorithm: projected gradient descent}
 \label{subsec: pgd}
 Inspired by \cite{cai2018spectral}, we design a projected gradient descent method for the problem \eqref{obj function}, which is summarized in  Algorithm \ref{alg: PGD-VHL}. 
 \begin{algorithm}[h]
 	\caption{PGD--VHL}
 	\label{alg: PGD-VHL}
 	\begin{algorithmic}
 		\State Input: $\calA, \vy, n, s, r$
 		\State Initialization:
 		\State $~\quad \hZ_0 = \hU_0 \hbSigma_0 {\hV_0}^\tranH=  \calP_r\calH\calAT(\vy) $
 		\State $~\quad \hL_0 =\hU_0 {\hbSigma_0}^{1/2},\quad \hR_0 = \hV_0 {\hbSigma_0}^{1/2}$
 		\State $~\quad \hM_0 = \begin{bmatrix}\hL_0^\tranH &\hR^\tranH_0 \end{bmatrix}^\tranH$
 		\State $~\quad (\mL_0, \mR_0) = \calP_{\calM}((\hL_0, \hR_0))$
 		\State $~\quad \mM_0 = \begin{bmatrix}\mL_0^\tranH &\mR^\tranH_0 \end{bmatrix}^\tranH$
 		
 		\While{ not convergence}
 		\State $\mM_{t+1} = \calP_{\calM}\left( \mM_t - \eta \nabla f(\mM_t) \right)$.
 		\EndWhile
 	\end{algorithmic}
 \end{algorithm}
 The initialization involves two steps: (1) computes the best rank $r$ approximation of $\calH\calAT(\vy)$ via one step hard thresholding $\calP_r(\cdot)$, where $\calAT$ is the adjoint of $\calA$  and $\calP_r(\mZ) $ is the best rank $r$ approximation of $\mZ$; (2) projects the low-rank factors of best rank-$r$ approximated matrix onto the convex feasible set $\calM$. 
 Given a matrix {\small $\mM = \begin{bmatrix}\mL^\tranH & \mR^\tranH\end{bmatrix}^\tranH$}, the projection onto $\calM$, denoted by {\small $\begin{bmatrix}\widehat{\mL}^\tranH & \widehat{\mR}^\tranH\end{bmatrix}^\tranH$}, has a closed form solution:
 \begin{align*}
 	\widehat{\mL}_j = \begin{cases}
 		\mL_j&\text{ if }\fronorm{\mL_j} \leq \sqrt{\frac{\mu r\sigma}{n}}\\
 		\frac{1}{\fronorm{\mL_j}}\mL_j\cdot \sqrt{\frac{\mu r\sigma}{n}} &\text{ otherwise}
 	\end{cases}
 \end{align*}
 for $0 \leq j \leq n_1-1$ and 
 \begin{align*}
 	\ve_k^\tran\widehat{\mR} = \begin{cases}
 		\ve_k^\tran \mR &\text{ if } \twonorm{\ve_k^\tran\mR} \leq \sqrt{\frac{\mu r\sigma}{n}}\\
 		\frac{\ve_k^\tran\mR}{\twonorm{\ve_k^\tran\mR}}\cdot \sqrt{\frac{\mu r\sigma}{n}} &\text{ otherwise}
 	\end{cases}
 \end{align*}
 for $0\leq k\leq n_2-1$.
 Let $\mM_t$ be the current estimator. The algorithm updates $\mM_t$ along gradient descent direction $-\nabla f(\mM_t) $ with step size $\eta$, followed by projection onto the set $\calM$.
 The gradient of $f(\mM)$ is computed with respect to Wirtinger calculus  given by 
 $\nabla f =\begin{bmatrix}\nabla^\tranH_{\mL} f& \nabla^\tranH_{\mR} f \end{bmatrix}^\tranH  $
 where  
 \begin{align*}
 	\nabla_{\mL} f= & \left(\calG\calAT\left(\calA\calGT(\mL\mR^\tranH) - \mD\vy\right)\right)\mR  +\lb(\calI - \calG\calGT)(\mL\mR^\tranH)\rb\mR + \frac{1}{4}\mL(\mL^\tranH\mL - \mR^\tranH\mR),\\
 	\nabla_{\mR} f = &\left(\calG\calAT\left(\calA\calGT(\mL\mR^\tranH) - \mD\vy\right)\right)^\tranH\mL + \lb(\calI - \calG\calGT)(\mL\mR^\tranH)\rb^\tranH\mL + \frac{1}{4}\mR(\mR^\tranH\mR - \mL^\tranH\mL).
 \end{align*}
 Indeed PGD--VHL algorithm can be efficiently implemented.  To obtain the computational cost of $\nabla f$, we first introduce some notations. Let $\calH_{v}$ be the Hankel operator which maps a vector $\vx\in\C^{n}$ into an $n_1\times n_2$ matrix,
 \begin{align*}
 	\calH_{v}(\vx) = \begin{bmatrix}
 		x_0 &\cdots &x_{n_2-1}\\
 		\vdots &\ddots &\vdots\\
 		x_{n_1-1}&\cdots &x_{n-1}\\
 	\end{bmatrix},
 \end{align*}
 where $x_i$ is the $i$-th entry of $\vx$.  The adjoint of $\calH_{v}$, denoted by $\calH_{v}^\ast$, is a linear mapping from $n_1\times n_2$ to $n$. It can be seen that $\calH^\ast_v\left( \mL_v\mR_v^\tranH\right)$ can be computed via $r$ fast convolutions by noting that 
 \begin{align*}
 	\left[\calH^\ast_v\left( \mL_v\mR_v^\tranH\right) \right]_i &= \left[\calH^\ast_v\left(\sum_{s=1}^r \mL_v[:,s]\overline{\mR_v}[:,s]^\tran\right) \right]_i\\
 	&=\sum_{s=1}^r\left[\calH^\ast_v\left( \mL_j[:,s]\bar{\mR}[:,s]^\tran\right) \right]_i\\
 	&=\sum_{s=1}^r \left( \mL_v[:,s] \ast \overline{\mR_v}[:,s]\right)[i],
 \end{align*}
 where $\mL_v\in\C^{n_1\times r}$ and $\mR_v\in\C^{n_2\times r}$.
 In addition, we can compute $\left(\calH_v(\vx)\right)\mR_v$  by $r$ fast Hankel matrix–vector multiplications, that is,
 \begin{align*}
 	\left(\calH_v(\vx)\mR_v\right)[j,s] 
 	&=\sum_{k=0}^{n_2-1}\vx[j+k]\mR_v[k,s]\\
 	&=\sum_{k=0}^{n_2-1}\tilde{\vx}[n-1 - j-k]\mR_v[k,s]\\
 	&= (\tilde{\vx} \ast \mR_v[:,s])[n-1-j],
 \end{align*}
 where $ \tilde{\vx}$ is a vector reversing the order of $\vx$. 
 Therefore the computational complexity of both $\calH_{v}^\ast(\mL_v\mR_v^\tranH)$ and $\left(\calH_v(\vx)\right)\mR_v$ is $\calO(rn\log n)$ flops. Moreover, the authors in \cite{chen2020vectorized} show that $\calH(\mX) = \mP\widetilde{\calH}(\mX)$, where $\widetilde{\calH}(\mX)$ is a matrix constructed by stacking all $\{\calH_{v}(\ve_\ell^\tran\mX)\}_{\ell=1}^s$ on top of one another, and $\mP$ is a permutation matrix. Therefore we can compute $\calGT(\mL\mR^\tranH)$ and $\calG\calD(\mX)\mR$ by using $\calO(srn\log n)$ flops. Thus the implementation of our algorithm is very efficient and the main computational complexity in each step is $\calO(sr^2n + srn\log n)$.
 
 \section{Main result}
 \label{sec: main result}
 In this section, we provide a theoretical analysis of PGD--VHL under a random subspace model. 
 
 \begin{assumption} 
 	\label{assumption 2}
 	The column vectors $\{\vb_i\}_{i=0}^{n-1}$ of $\mB^\tranH$ are independently and identically drawn from a distribution $F$ which satisfies the following conditions:
 	\begin{align}
 		\label{eq: isotropy property}
 		\E{\vb_i\vb_i^\tranH} &= \mI_s, \quad i=0,\cdots, n-1,\\
 		\label{eq:incoherence}
 		\max_{0\leq \ell\leq s-1}  | \vb_i[\ell]|^2 &\leq \mu_0, \quad i=0,\cdots, n-1.
 	\end{align}
 \end{assumption}
 \begin{remark}
 	{\color{black} Assumption \ref{assumption 2} is a standard assumption in RIPless compressed sensing \cite{candes2011probabilistic} and blind super-resolution \cite{chi2016guaranteed, yang2016super, li2019atomic, suliman2018blind,chen2020vectorized}. It implies the spectral flatness over point spread functions, which is satisfied in OFDM signals \cite{barbieri2010ofdm} and noisy radar waveforms \cite{pralon2021analysis}.  
 	} Assumption \ref{assumption 2} holds with $\mu_0 = 1$ by many common random ensembles, for instance, when the components of $\vb$ are Rademacher random variables taking the values $\pm 1$ with equal probability or when $\vb$ is uniformly sampled from the rows of a Discrete Fourier Transform (DFT) matrix. 
 	
 \end{remark}
 
 Now we present the main result, whose proofs are deferred to Section \ref{sec: proof}.
 \begin{theorem}
 	\label{main result}
 	Let $\mu \geq \mu_1$ and $\sigma = \sigma_1({\widehat\bSigma}_0)/(1-\varepsilon)$ for $0 \leq \varepsilon \leq 1/3$. Let $\eta\leq \frac{\sigma_r}{9000(\mu_0 \mu sr\sigma_1)^2}$, $\beta = \frac{\sigma_r}{72}$, and {\small$\mM^\natural = \begin{bmatrix}{\mL^\natural}^\tranH\quad{\mR^\natural}^\tranH\end{bmatrix}^\tranH$}. Suppose $\mX^\natural$ obeys the Assumption \ref{assumption 1} and the subspace $\mB$ satisfies the Assumption \ref{assumption 2}. If 
 	\begin{align*}
 		n\geq c_0 \varepsilon^{-2}\mu_0^2\mu s^2r^2 \kappa^2\log^2(sn),
 	\end{align*}
 	with probability at least $1-c_1 (sn)^{-c_2}$, the sequence $\{\mM_t\}$ returned by Algorithm \ref{alg: PGD-VHL} satisfies
 	\begin{align}
 		\label{ineq: convergence rate}
 		\dist^2(\mM_t,\mM^\natural)\leq (1-\eta\beta)^t\cdot \frac{\varepsilon^2\sigma_r}{\mu_0 s},
 	\end{align}
 	where $c_0, c_1, c_2$ are absolute constants, $\sigma_1 = \sigma_1(\mZ^\natural)$, $\sigma_r = \sigma_r(\mZ^\natural)$, $\kappa$ is the condition number of $\mZ^\natural$, and the distance $\dist(\mM,\mM^\natural)$ is defined as
 	\begin{align*}
 		\dist(\mM,\mM^\natural) = \min_{\mQ\mQ^\tranH = \mQ^\tranH\mQ = \mI_r} \fronorm{\mM - \mM^\natural \mQ}.
 	\end{align*}
 \end{theorem}
 {\color{black}
 \begin{remark}
    It is worth noting that our results require slightly milder assumptions than that in \cite{chen2020vectorized}. The theoretical performance in \cite{chen2020vectorized} is established based on an additional assumption, which requires a lower bound of $\ell_2$ norm of the row vector of $\mB$. However, the performance guarantee of PGD--VHL is independent of this assumption.
 \end{remark}
 }

 \begin{remark}
 \label{remark: bound}
 	Compared with the sample complexity established in \cite{chen2020vectorized} for the nuclear norm minimization method, which is $n \geq c\mu_0 \mu_1 \cdot sr\log^4(sn)$, Theorem \ref{main result} implies that PGD--VHL is sub-optimal in terms of $s$ and $r$. {\color{black}{The extra $r$ factor is caused by the technical derivation. Since the convergence is established based on the Frobenius norm and the Frobenius norm of the initial error is bounded by its spectral norm. Details can be found in Section~\ref{sec: init}}. The extra $s$ factor is introduced to ensure the initial error to be sufficiently small, i.e., on the order of $1/(\mu_0 s)$, which helps to derive the linear convergence rate of PGD--VHL. We admit that it is an artifact of our proof because} subsequent numerical experiments indicate that there approximately exists a linear relationship between $n$ and $s$ or $n$ and $r$.
 	 
 \end{remark}
 
 \begin{remark}
 	Theorem \ref{main result} implies that PGD--VHL converges to $\mM^\natural$ with a linear rate. Therefore, after $T=\calO((\mu_0\mu sr\kappa)^2\log(1/\epsilon))$ iterations, we have $\dist^2(\mM_T, \mM^\natural) \leq \epsilon\cdot \dist^2(\mM_0, \mM^\natural) $. Given the iterates $\mM_T$ returned by PGD--VHL, we can estimate $\mX_T$ by $\calD^{-1}\calGT(\mL_T \mR_T^\tranH)$. 
 \end{remark}
 \begin{remark}
 	As already mentioned, once the data matrix $\mX^\natural$ is recovered, the locations $\{ \tau_k\}_{k=1}^r$ can be computed from it by spatial smoothing MUSIC algorithm \cite{evans1981high,evans1982application,yang2019source,chen2020vectorized} and the weights $\{ d_k , \vh_k \}_{ k=1 }^r$ can be estimated by solving an overdetermined linear system \cite{chen2020vectorized}.
 \end{remark}
{\color{black}
\begin{remark}
	Though we have mainly focused on the one-dimensional (1D) blind super-resolution problem, the model and analysis are also applicable for higher dimensional case. Due to space limitations, we omit the theoretical results but provide numerical simulations for two-dimensional (2D) blind super-resolution problem in Section V.B.
\end{remark}
}

\begin{figure}[htb]
	\centering
	
	\subfigure{
		\begin{minipage}[t]{0.3\linewidth}
			\centering
			\includegraphics[width=5.15cm]{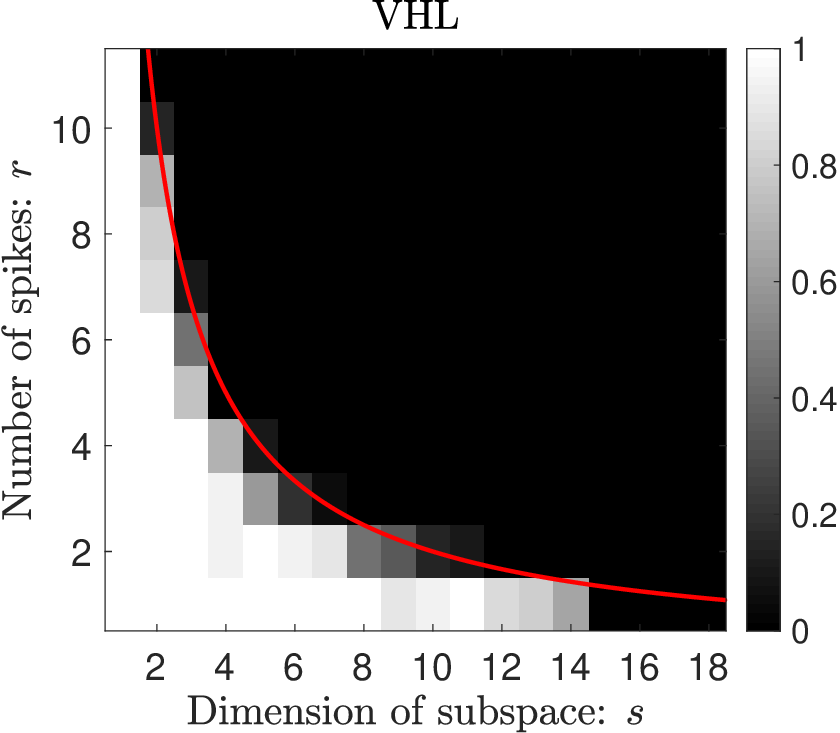}
			\centerline{(a)}\medskip
		\end{minipage}
	}
	\subfigure{
		\begin{minipage}[t]{0.3\linewidth}
			\centering
			\includegraphics[width=5.15cm]{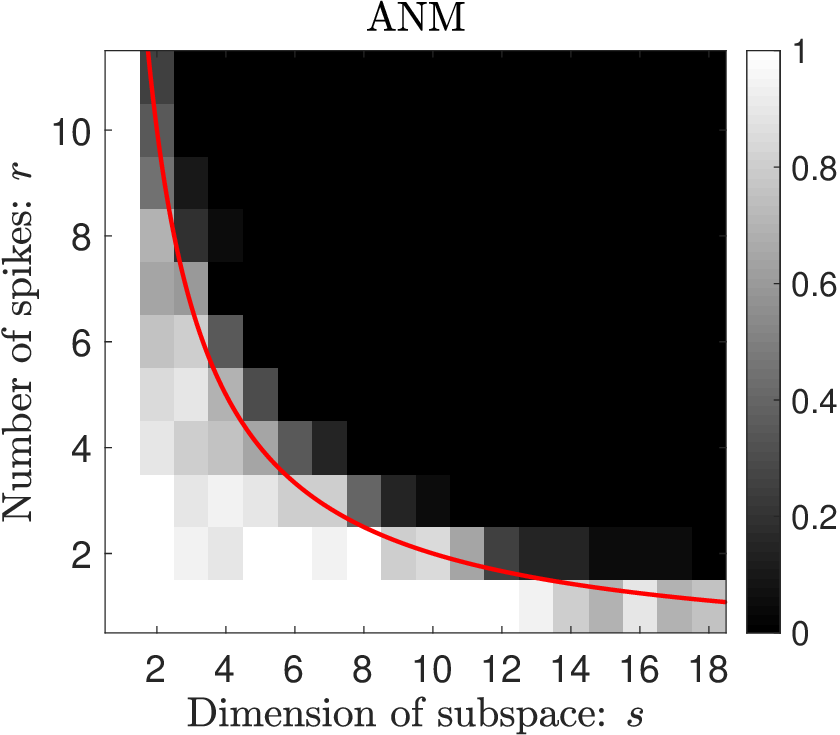}
			\centerline{(b)}\medskip
		\end{minipage}
	}
	\subfigure{
		\begin{minipage}[t]{0.3\linewidth}
			\centering
			\includegraphics[width=5cm]{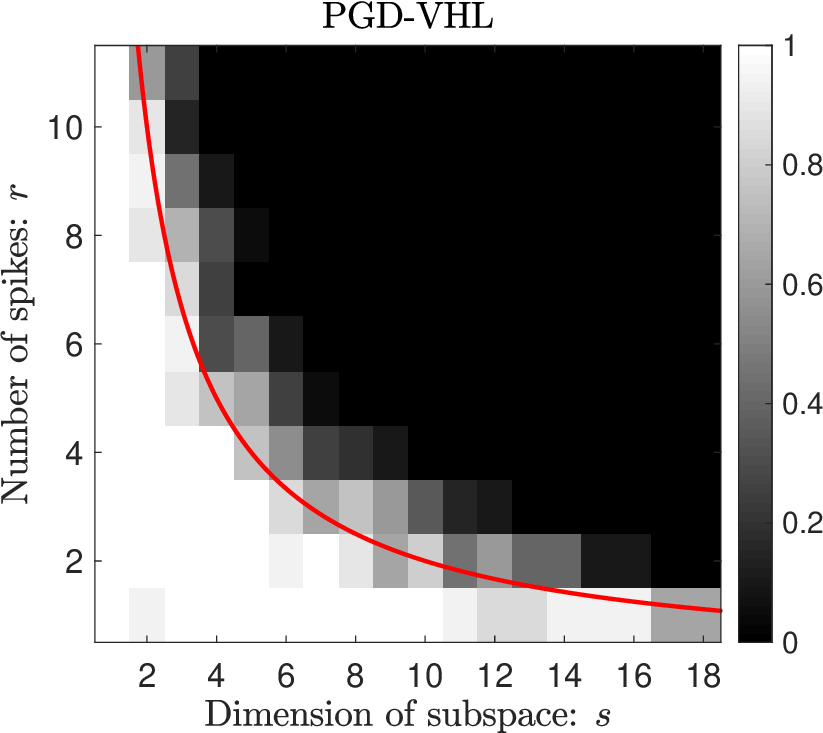}
			\centerline{(c)}\medskip
		\end{minipage}
	}
	
	\subfigure{
		\begin{minipage}[t]{0.3\linewidth}
			\centering
			\includegraphics[width=5.0cm]{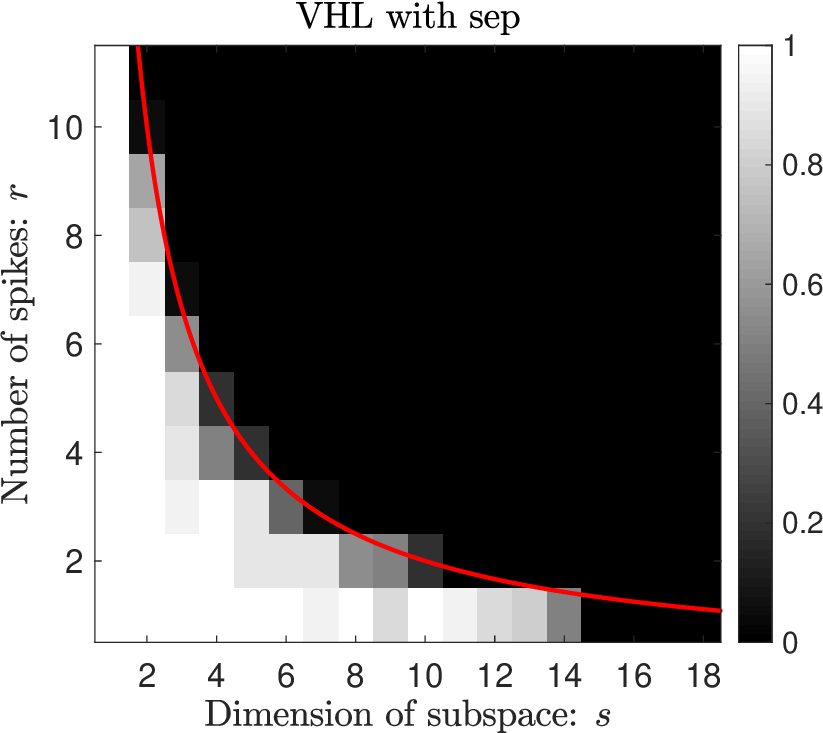}
			\centerline{(d)}\medskip
		\end{minipage}
	}
	\subfigure{
		\begin{minipage}[t]{0.3\linewidth}
			\centering
			\includegraphics[width=5.0cm]{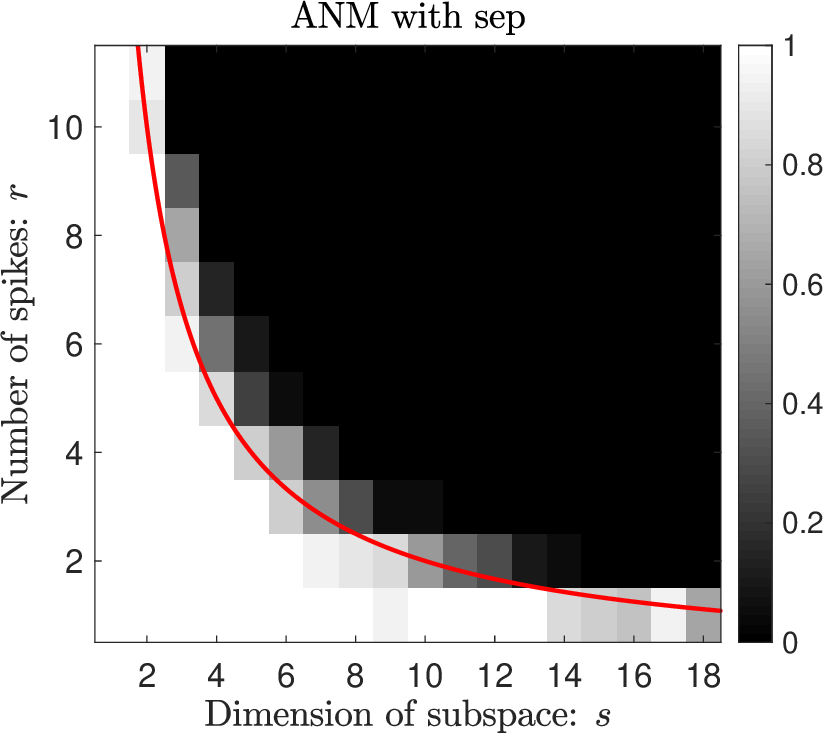}
			\centerline{(e)}\medskip
		\end{minipage}
	}
	\subfigure{
		\begin{minipage}[t]{0.3\linewidth}
			\centering
			\includegraphics[width=5.0cm]{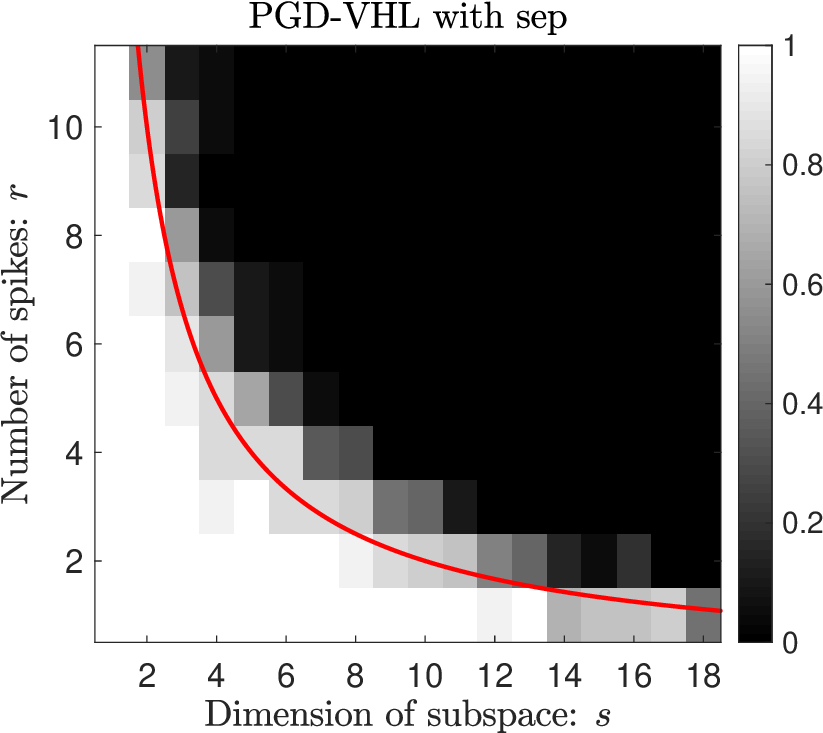}
			\centerline{(f)}\medskip
		\end{minipage}
	}
	
	\caption{The phase transitions of VHL, ANM and PGD-VHL when $n=64$. Top: frequencies are randomly generated; Bottom: frequencies obey the separation condition $\Delta:= \min_{k\neq j}\lab\tau_k - \tau_j\rab\geq 1/n$. The red curve plots the hyperbola curve $rs = 20$.}
	\label{fig:phase transition}
\end{figure}

\section{Numerical simulations}
\label{sec: numerical}
{\color{black} In this section, a series of numerical results are provided to illustrate the performance of PGD--VHL. We conduct simulations on 1D signals and 2D signals separately. Moreover, we implement our algorithm with 2D MUSIC \cite{ berger2010signal, zheng2017super} to solve the joint delay-Doppler estimation problem, which is an important issue arising in orthogonal frequency-division multiplexing (OFDM) signals.  The numerical simulations are executed from MATLAB R2021b on a macOS machine with  multi-core Intel CPU  at 2.3 GHz CPU and 16 GB RAM. Our code is available at ``https://github.com/jcchen2017/PGDVHL".

\subsection{Simulation for 1D Signals} 
We begin by providing the numerical results for 1D signals. The  data matrix $\mX^\natural\in\C^{s\times n}$ is generated by $\sum_{k=1}^r d_k\vh_k\va_{\tau_k}^\tran$. Here the locations $\{\tau_k\}_{k=1}^r$ of the point sources are randomly generated from $[0,1)$, the coefficients $\{\vh_k\}_{k=1}^r$ are i.i.d. sampled from standard Gaussian with normalization, and the amplitudes $\{d_k\}_{k=1}^r$ are selected to be $d_k = (1+10^{c_k})e^{-\imath \phi_k}$, where $c_k$ is uniformly sampled from $[0,1]$ and $\phi_k$ is uniformly sampled from $[0,2\pi)$.  Moreover, the columns of  $\mB$ are uniformly sampled from the DFT matrix. The stepsize of PGD--VHL is chosen via backtracking line search. 

}

 The first experiment studies the recovery ability of PGD--VHL through the framework of phase transition and we compare it with two convex recovery methods: VHL \cite{chen2020vectorized} and ANM \cite{yang2016super}. Both VHL and ANM are solved by CVX \cite{cvx}. PGD--VHL will be terminated if $\twonorm{y-\calA(\mX_t)}\leq 10^{-5}$ or a maximum number of iterations is reached. The tests are conducted with $n=64$ and the varied $s$ and $r$. We repeat 20 random trials and record the probability of successful recovery in our tests. A trial is declared to be successful if $\fronorm{\mX_t - \mX^\natural}/\fronorm{\mX^\natural} \leq 10^{-3}$. Figure \ref{fig:phase transition}(a), \ref{fig:phase transition}(b) and \ref{fig:phase transition}(c) show the phase transitions of VHL, ANM and PGD--VHL when the locations of point sources are randomly generated, and Figure \ref{fig:phase transition}(d), \ref{fig:phase transition}(e) and \ref{fig:phase transition}(f) illustrate the phase transitions of VHL, ANM and PGD--VHL when the separation condition $\Delta:=\min_{j\neq k}|\tau_j -\tau_k| \geq 1/n $ is imposed. In this figure, white color means successful recovery while black color indicates failure. It is interesting to observe that PGD--VHL has a higher phase transition curve than VHL whether the separation condition is satisfied or not. Moreover, by comparing Figure \ref{fig:phase transition}(b) with (c), we observe that PGD--VHL is less sensitive to the separation condition than ANM.


In the second experiment, we study the phase transition of PGD--VHL when one of $r$ and $s$ is fixed. Note that in this test, the separation condition is not imposed for PGD--VHL. Figure 2(a) indicates an approximately linear relationship between $s$ and $n$ for the successful recovery when the number of point sources is fixed to be $r = 4$. The same linear relationship between $r$ and $n$ can be observed when the dimension of the subspace is fixed to be $s = 4$, see Figure 2(b). Therefore there exists a gap between our theory and empirical observation and we leave it as future work.

\begin{figure}[htb]
	\centering
	\subfigure{
		\centering
		\includegraphics[width=.3\textwidth]{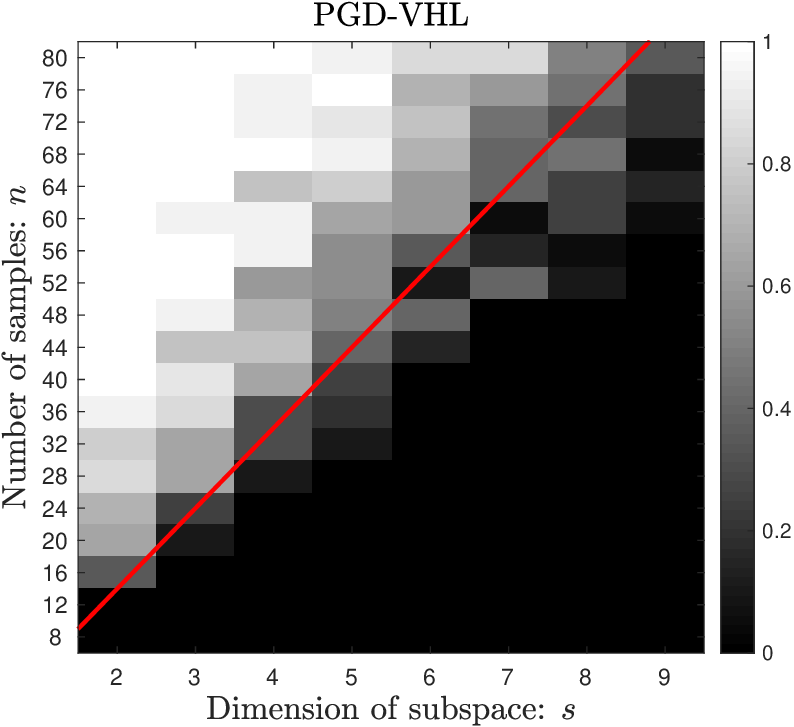}
	}
	\hspace{0.4cm}
	\subfigure{
		\centering
		\includegraphics[width=.3\textwidth]{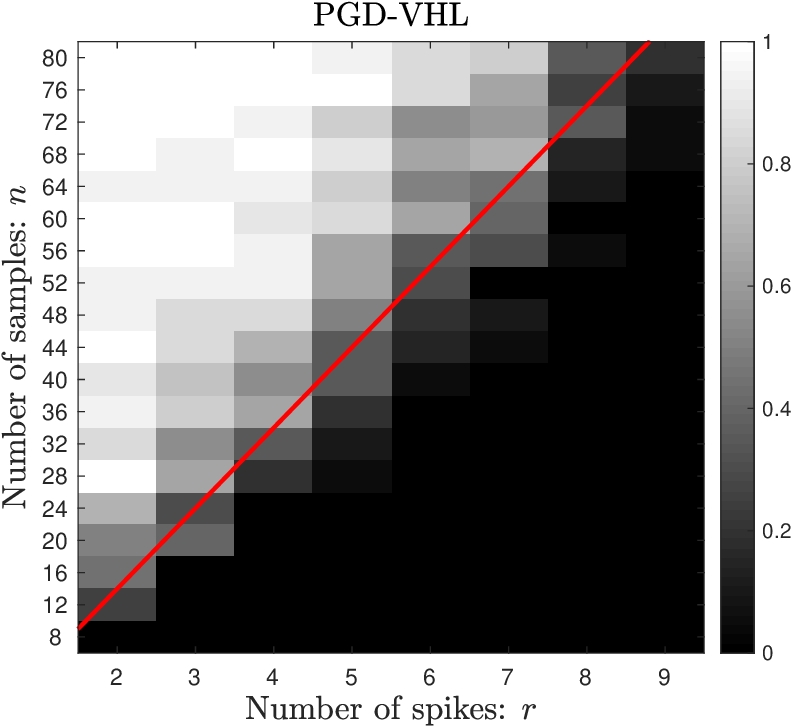}
	}
	
	\caption{(a) The phase transition of PGD--VHL for varying $n$ and $s$ when $r = 4$. The red line plots the straight line $n = 2.5s$. (b) The phase transition of PGD--VHL for varying $n$ and $r$ when $s = 4$. The red line plots the straight line $n = 2.5r$.}
	\label{fig: phase transition n}
\end{figure}

{\color{black}
In the third simulation, we investigate the convergence rate of PGD--VHL for $n = 1024$ with fixed $s$ or $r$. The results are shown in Figure~\ref{fig: convergence}. The $y$-axis  denotes $\log\left(\fronorm{\mX_t - \mX^\natural}/\fronorm{\mX^\natural}\right)$ 
and the $x$-axis represents the iteration number. It can be clearly seen that PGD--VHL converges linearly as shown in our main theorem. Also it is worth pointing out that PGD--VHL can be implemented in high dimensional regimes, where we take $n=1024$. 

\begin{figure}[htb]
	\centering
	\subfigure{
		\centering
		\includegraphics[width=.3\textwidth]{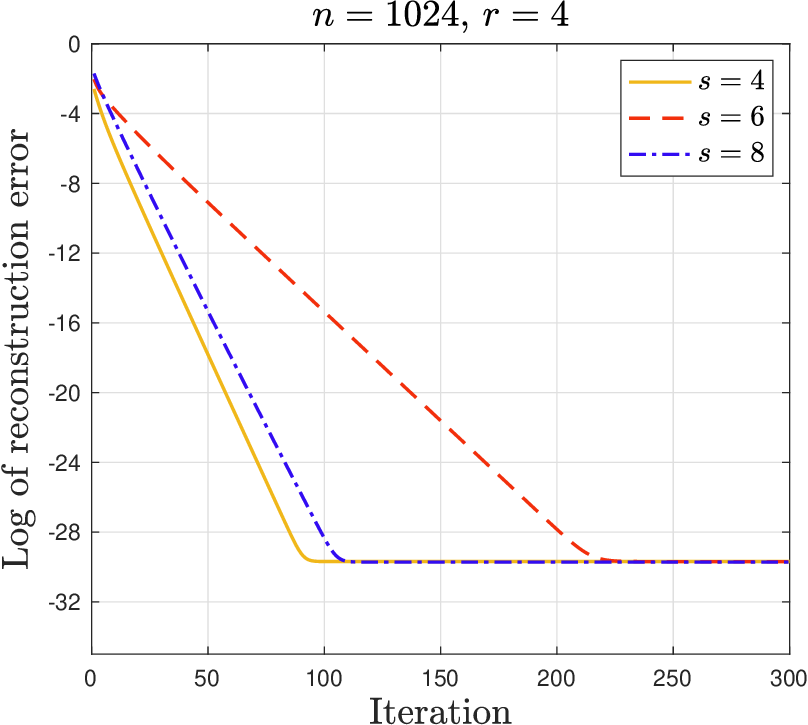}
	}
	\hspace{0.4cm}
	\subfigure{
		\centering
		\includegraphics[width=.3\textwidth]{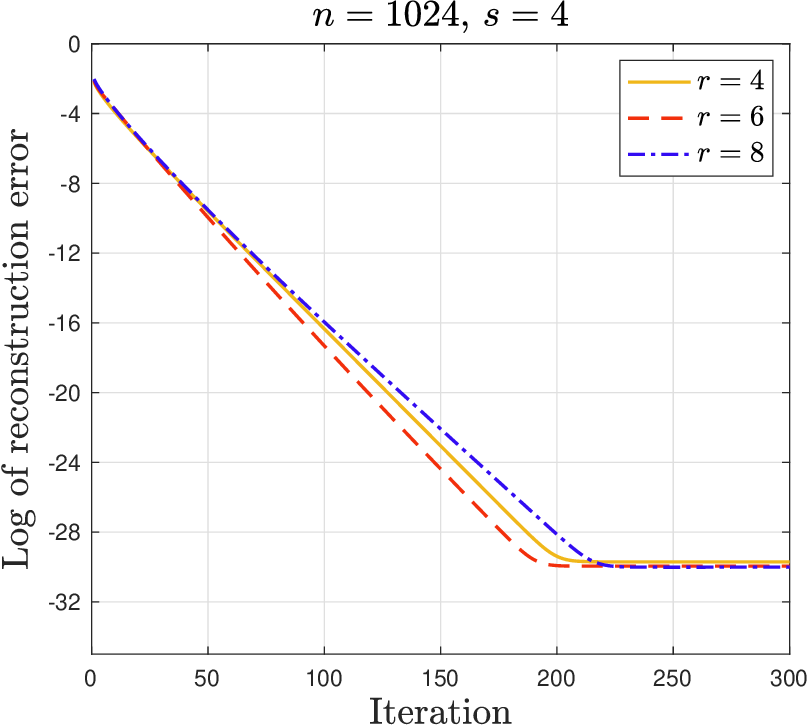}
	}
	
	\caption{{\color{black}(a) Convergence of PGD--VHL for varying $s=4, 6, 8$ when $n = 1024 $ and $r=4$.  (b) Convergence of PGD--VHL for varying $r=4, 6, 8$ when $n=1024$ and $s=4$.}}
	\label{fig: convergence}
\end{figure}
}

\begin{figure}[htb]
	\centering
	\includegraphics[width=.3\textwidth]{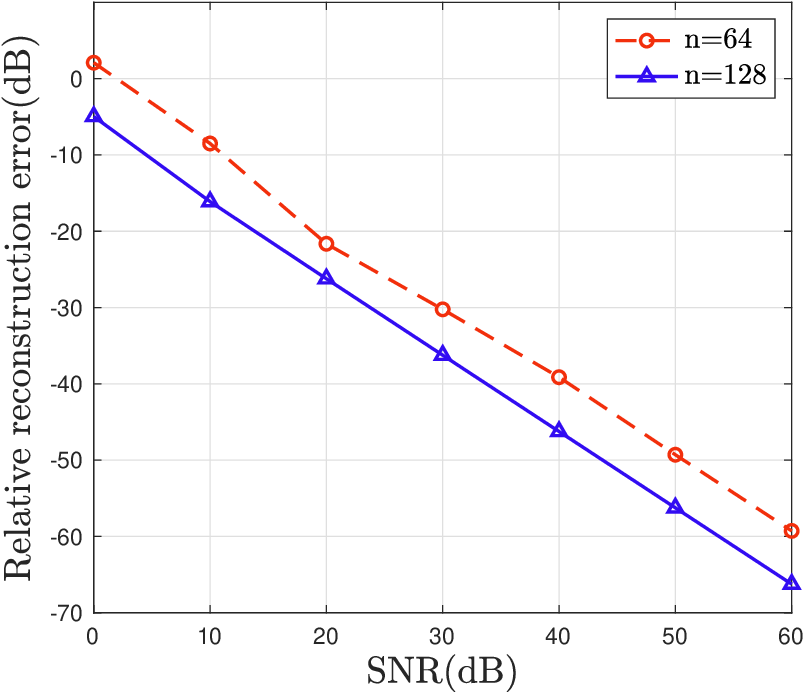}
	\caption{Performance of PGD--VHL under different noise levels}
	\label{fig: noise}
\end{figure}

In the fourth simulation, we  conduct the tests to demonstrate the robustness of PGD--VHL to additive noise. More specifically, we collect the measurements corrupted by the noise vector $\ve = \sigma_{\ve}\cdot \vecnorm{\vy}{2} \cdot   \vw/\vecnorm{\vw}{2}$,
where $\vy$ is the uncontaminated observations, $\vw$ is the standard Gaussian vector with i.i.d. entries and $\sigma_{\ve}$ denotes the noise level. In the tests, the noise level $\sigma_{\ve}$ is taken from $10^{-3}$ to $1$, corresponding to the signal-to-noise ratio (SNR) from $60$ to $0$ dB. For each $\sigma_{\ve}$, 10 random trials are conducted with $s=r=4$. As for the number of measurements, we choose $n=64$ and $n=128$ for comparison. PGD-VHL is set to be terminated when $\fronorm{\mX^{t+1} - \mX^{t}}/\fronorm{\mX^t}\leq 10^{-7}$. In Figure~\ref{fig: noise}, the average relative reconstruction error is plotted with SNR. It can be clearly seen that the relationship between the relative reconstruction error and the noise level is linear for PGD--VHL. Moreover, the relative reconstruction error decreases with the increase of the number of measurements.

{\color{black} We finally compare the running time for ANM, VHL and PGD--VHL when the number of measurements is varied, the number of spikes $r$ is fixed to be $3$ and the dimension of subspace $s$ is also fixed to be $3$. Note that both VHL and ANM are solved by SDPT3 \cite{toh1999sdpt3} based on CVX \cite{cvx}. We repeat
10 random trials for each test. The average computational time for each tested algorithms are shown in Table I. The  symbol $``-"$  indicates that the algorithm was terminated due to the lack of memory. It can be seen that PGD--VHL significantly improve the running time compared with ANM and VHL when $n$ is large.


\begin{table}[h!]
\centering
\caption{Running time comparison for 1D signals when $s=r=3$.}
\label{tab1: running time}
\setlength{\tabcolsep}{3mm}{
\begin{tabular}{ c c c c c } 

 \hline
 Methods & $n=64$ & $n=128$ & $n=256$ &$n=512$ \\ [0.5ex] 
 \hline
 ANM & 1.6278s &  7.2992s& 67.6007s &$-$\\ 

 VHL & 62.2369s &  748.3695s& $-$ &$-$ \\

 PGD--VHL & 1.3254s &  4.4417s& 19.3004s & 57.4518s\\ [1ex]
 \hline 
\end{tabular}}
\end{table}

}



{\color{black}
	\subsection{Simulation for 2D Signals}
	In this part, we evaluate the performance of our algorithm for 2D signals. The data matrix is given by $\mX^\natural = \sum_{k=1}^{r}d_k \vh_k  ( \va_{\tau_{2k}}\otimes \va_{\tau_{1k}} )^\tran\in\C^{s\times n_1n_2}$ and the samples are generated by 
	\begin{align}
		\label{eq: 2D sample}
		\vy[j] = \la\vb_j\ve_j^\tran, \mX^\natural \ra, \quad j=0,\cdots, n_1n_2-1.
	\end{align}
	Here the two dimensional locations $\bfm{\tau}_k:=(\tau_{1k}, \tau_{2k})$ are uniformly sampled from $[0,1)\times [0,1)$, the amplitudes $\{d_k\}_{k=1}^r$, the coefficients $\{\vh_k\}_{k=1}^r$ and the subspace columns $\{\vb_j\}_{j=0}^{n_1n_2-1}$ are generated by the same way as the 1D case. Let $\mX^{\natural}_{\ell} = \sum_{k=1}^{r}d_k e^{-2\imath \pi \tau_{2k} \cdot \ell} (\vh_k\va_{\tau_{1k}}^\tran)$ be an $s\times n_1$ matrix, where $\ell =0,\cdots,n_2-1$. Let $\calH(\mX^\natural)$ be the two-fold vectorized Hankel matrix of $\mX^\natural$ defined as follows:
	\begin{align*}
		\calH(\mX^\natural) = \begin{bmatrix*}
			\calH(\mX_0^\natural) &\cdots &\calH(\mX_{K_2-1}^\natural)\\
			\vdots &\ddots &\vdots\\
			\calH(\mX_{K_1-1}^\natural) &\cdots &\calH(\mX_{n_2}^\natural)\\
		\end{bmatrix*}\in\C^{sL_1K_1 \times L_2 K_2},
	\end{align*}
	where $\calH(\mX_\ell^\natural)\in\C^{sL_1\times L_2}$ is the vectorized Hankel matrix defined in \eqref{vhl}. Here $L_1 + L_2 = n_1+1$ and $K_1 + K_2 = n_2+1$. It has been shown in \cite{chen2020vectorized} that $\calH(\mX^\natural)$ is a rank-$r$ matrix. Therefore, we can naturally generalize our algorithm to the 2D case, and then use the two-dimensional PGD--VHL (PGD--VHL (2D)) to recover $\mX^\natural$ from \eqref{eq: 2D sample}. The phase transition is shown in Figure~\ref{fig: phase transition pgd-vhl 2D} and the convergence rate of PGD--VHL (2D) for different $n_1$ and $n_2$ is shown in Figure \ref{fig: convergence 2D}. Overall, the performance of PGD--VHL (2D) exhibits a similar phenomenon to the 1D case.
	
	\begin{figure}[htb]
		\centering
		\includegraphics[width=.3\textwidth]{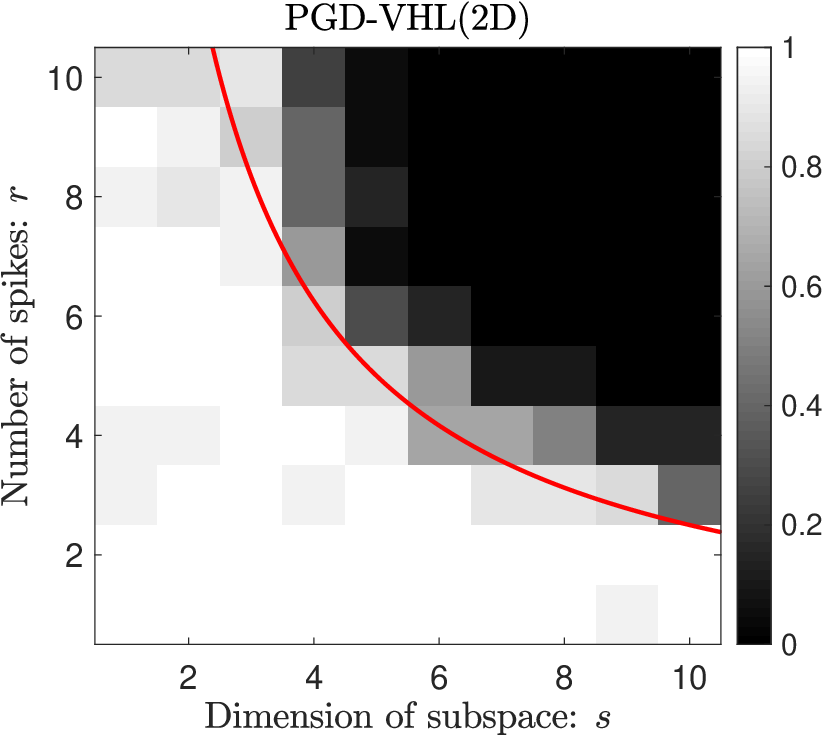}
		\caption{{\color{black} The phase transition of PGD--VHL (2D) for varying $s$ and $r$ when $n_1 = 13$ and $n_2= 9 $. The locations are randomly generated. The red line plots the straight line $sr = 25$. }}
		\label{fig: phase transition pgd-vhl 2D}
	\end{figure}
	
	\begin{figure}[htb]
		\centering
		\includegraphics[width=.3\textwidth]{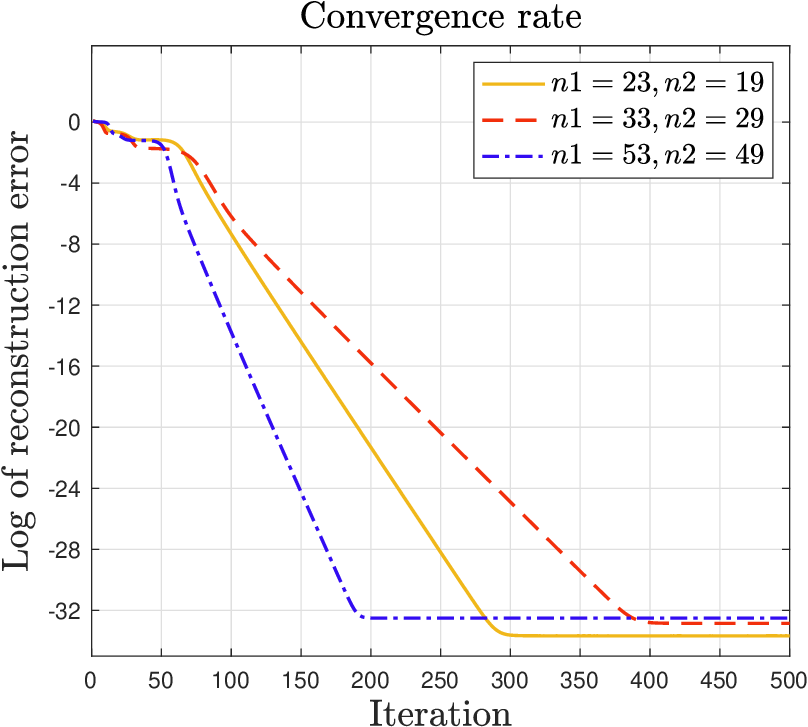}
		\caption{{\color{black}Convergence of PGD--VHL (2D) for 2D signals with $(n_1, n_2)\in \{ (23, 19), (33, 29), (53, 49) \}$. Here we fix $s=3$ and $r=3$. }}
		\label{fig: convergence 2D}
	\end{figure}
	
	\subsection{Simulation for joint delay-Doppler estimation from OFDM signals}
	Furthermore, we evaluate the performance of PGD--VHL (2D) for the problem of joint delay-Doppler estimation from OFDM signals. In this problem, the transmitted signal is divided into $M$ blocks and $N$ orthogonal subcarriers are used in each block. Then the received samples in the $n$-th subcarrier and $m$-th block can be formulated as \cite{zheng2017super}
	\begin{align}
		\label{ofdm model}
		\vy_m[n] = \sum_{k=1}^{r} d_k e^{-2\imath \pi (n\cdot \Delta f\tau_k + m \cdot \bar{T}f_k  ) } \vg_m[n],
	\end{align}
	where $r$ is the number of propagation paths for communication channel, $\{\tau_k, f_k\}_{k=1}^r$ are the delays and Doppler frequencies, $\{d_k\}_{k=1}^r\subset \C$ denote the channel coefficients, $\Delta f$ is the frequency spacing of adjacent subcarriers, $\bar{T}$ is the duaration of each transmission block with $\bar{T}f_k \ll 1$ as stated in \cite{zheng2017super}, and $\{\vg_m[n]\}_{n=0,\cdots, N-1}$ denote data symbols in the $m$-th block. For sake of simplicity, we define $\phi_k = \Delta f\tau_k \in [0,1)$, $\psi_k = \bar{T}f_k\in[0,1)$.  Concatenating $\vy_m[n]$ in the vector $\vy_m\in\C^N$ and stacking $\vy_m$ yield that 
	\begin{align*}
		\vy  = \left(\sum_{k=0}^{r} d_k \va_{\psi_k} \otimes \va_{\phi_k} \right) \odot \vg \in\C^{MN},
	\end{align*}
	where $\vg = \begin{bmatrix}
		\vg_1 &\cdots &\vg_M
	\end{bmatrix}^\tran\in\C^{MN}$ and $\vg_m = \begin{bmatrix}
		\vg_m[0] &\cdots \vg_m[N-1]
	\end{bmatrix}^\tran\in\C^N$. As pointed in \cite{barbieri2010ofdm, vargas2022joint}, since the waveforms in OFDM have a flat spectrum, $\vg$ can be approximately represented as $\vg = \mB\vh$, where $\mB$ should obey the isotropy and incoherence properties in Assumption \ref{assumption 2}. 
	Then the received samples can be rewritten as
\begin{align}
	\label{real data model}
	\vy[j] = \la \vb_j \ve_j^\tran, \sum_{k=1}^{r} d_k \vh (\va_{\psi_k} \otimes  \va_{\phi_k})^\tran  \ra
\end{align}
for $j=0,\cdots, MN-1$. It can be seen that \eqref{real data model} is a special case of \eqref{eq: 2D sample}  where $\vh$ is independent to the two dimensional locations $\{(\psi_k, \phi_k)\}_{k=1}^r$. 

In our numerical simulation, We set $N = 13, M = 9$ and $s=r=4$. Each row of $\mB$ is generated from the following distribution described in  \cite{chi2016guaranteed,vargas2022joint}, i.e.,
\begin{align*}
    \vb_k = \begin{bmatrix}
    1 & e^{2\pi \imath f_k} &\cdots & e^{2\pi \imath (s-1)f_k}
    \end{bmatrix}^\tran
\end{align*}
for $k=0,\ldots, MN-1$, where $f_k$ is chosen uniformly at random in $[0,1]$. 
The locations $\{(\psi_k,\phi_k)\}_{k=1}^r$ are chosen uniformly at random from $[0,1)\times [0,1)$, and the coefficient vector $\vh$ is generated from standard Gaussian with normalization.
The data matrix $\sum_{k=1}^{r} d_k \vh (\va_{\psi_k} \otimes  \va_{\phi_k})^\tran $ can be firstly recovered via PGD--VHL (2D), then the locations $\{(\psi_k, \phi_k)\}_{k=1}^r$ are retrieved by 2D MUSIC and the channel coefficients $\{ d_k\}_{ k=1 }^r$ are estimated by solving an overdetermined linear system \cite{chen2020vectorized}. The results are presented in Figure~\ref{fig: 2D MUSIC}. It is shown that implementing PGD--VHL (2D) with 2D MUSIC can exactly recover delays and Doppler frequencies. 
\begin{figure}[htb]
	\centering
	\includegraphics[width=.35\textwidth]{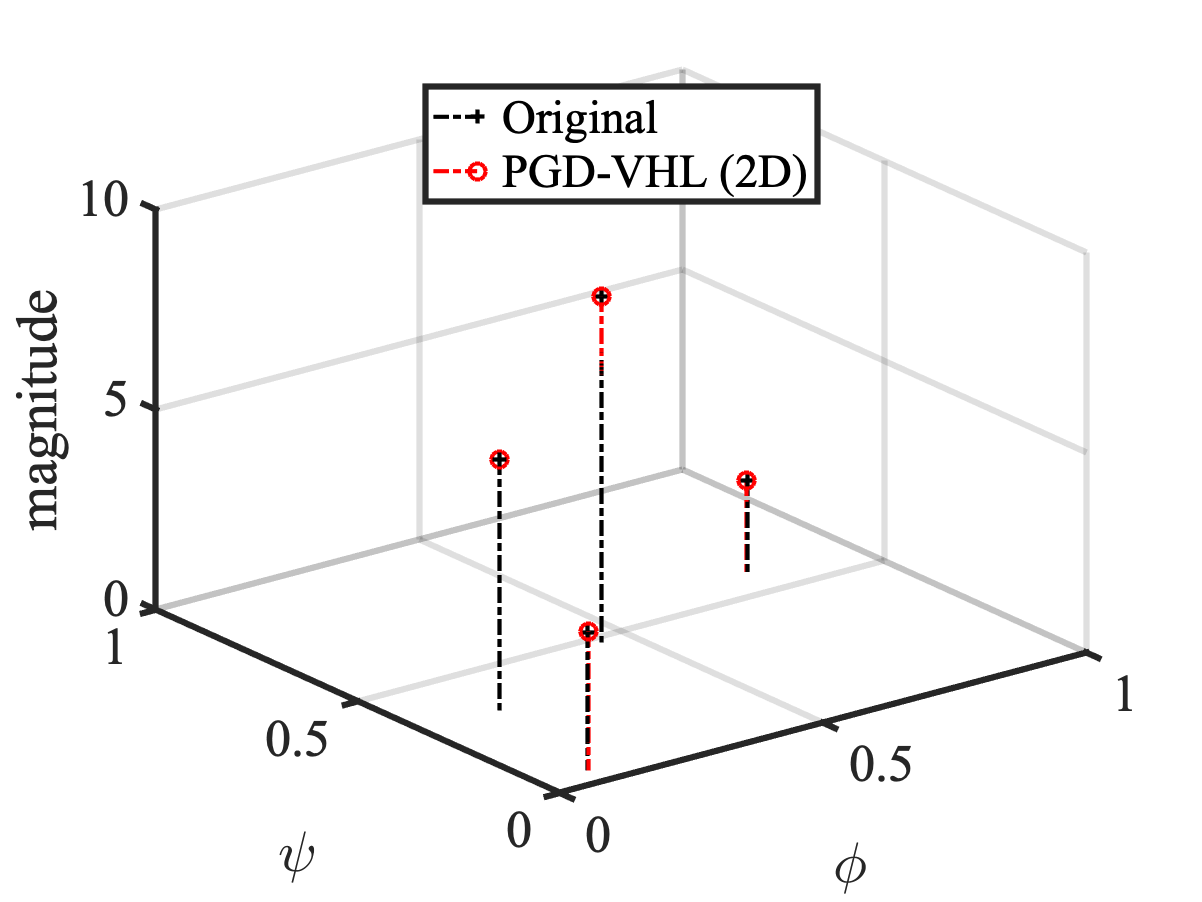}
	\caption{{\color{black} Performance of PGD-VHL (2D) for joint delay-Doppler estimation problem.}}
	\label{fig: 2D MUSIC}
\end{figure}	
	
%

}

\section{Proof of Theorem \ref{main result}}
\label{sec: proof}
The proof follows a well established route that has been widely used in non-convex optimization for low rank matrix recovery \cite{candes2015phase, zheng2016convergence, cai2018spectral}.  In a nutshell, the initialization provided in Algorithm \ref{alg: PGD-VHL} will be shown to lie in a basin of attraction where the sequence returned by Algorithm \ref{alg: PGD-VHL} converges linearly to the true solution. 
Despite this, the proof details are quite involved and substantially different. 
We first list two useful lemmas, whose proofs are deferred to Section \ref{sec: init} and \ref{proof local convergence}.
\begin{lemma}
	\label{lemma: initalization error}
	Suppose $\mZ^\natural$ is $\mu_1$-incoherent and $n\geq c_0\varepsilon^{-2}\kappa^2 \mu_0^2 \mu s^2 r^2 \log^2(sn)$. Then one has 
	\begin{align*}
		\dist^2(\mM_0 , \mM^\natural) \leq \frac{\varepsilon^2\sigma_r}{\mu_0s} 
	\end{align*}
	with probability at least $1-(sn)^{-c_1}$.
\end{lemma}

\begin{lemma}
	\label{lem: local convergence}
	Let $\{\mM_t\}$ be the sequence returned by Algorithm \ref{alg: PGD-VHL}. Denote $\bDelta_t := \mM_t - \mM^\natural\mQ_t$, where $$\mQ_t = \arg\min\limits_{\mQ\mQ^\tranH = \mQ^\tranH\mQ = \mI_r}\fronorm{\mM_t - \mM^\natural \mQ}.$$ Let $\eta\leq \frac{\sigma_r}{9000(\mu_0 \mu sr\sigma_1)^2}$, $\beta = \frac{\sigma_r}{72}$. Then with probability $1-(sn)^{-c}$,  one has 
	\begin{align*}
		\dist^2(\mM_{t}, \mM^\natural) \leq (1-\eta\beta)^{t} \dist^2(\mM_0,\mM^\natural) 
	\end{align*}
	for $t=1,2,\cdots$.
\end{lemma}
Combining Lemma \ref{lemma: initalization error} and Lemma \ref{lem: local convergence} together, we complete the proof. 

\subsection{Proof of Lemma \ref{lemma: initalization error}}
\label{sec: init}
We begin our presentation of the proof with a useful lemma whose proof is provided in Section \ref{proof lemma init opnorm}.
\begin{lemma}
	\label{lem: init opnorm}
	Suppose that $\mZ^\natural$ is $\mu_1$-incoherent. Then with probability at least $1-(sn)^{-c_1}$, the matrix $\hZ_0 = \calP_r(\calG\calD\calAT(\vy))$ obeys
	\begin{align}
		\label{ineq: deviation of init}
		\opnorm{\hZ_0 - \mZ^\natural}\leq c_0\sigma_1 \sqrt{\frac{\mu_0\mu_1 sr\log^2(sn)}{n}},
	\end{align} 
	where $c_0$ and $c_1$ are absolute constants.
\end{lemma}
By Lemma \ref{lem: init opnorm} and the assumption that $n$ should be larger than $c_0\varepsilon^{-2}\kappa^2 \mu_0^2 \mu s^2 r^2 \log^2(sn)$, the event
\begin{align*}
	\opnorm{\hZ_0 - \mZ^\natural}\leq c_1 \sqrt{\frac{\mu_0\mu_1 sr\log^2(sn)}{n}}\sigma_1(\mZ^\natural)\leq \varepsilon\sigma_1(\mZ^\natural)
\end{align*}
occurs with probability at least $1-(sn)^{-c_1}$, which implies that
\begin{align*}
	\sigma_1(\mZ^\natural) \leq \frac{\sigma_1(\hZ_0)}{1-\varepsilon} =: \sigma \text{ and } \sigma_1(\hZ_0) \leq  (1+\varepsilon) \sigma_1 \leq 2\sigma.
\end{align*}
By the definition of $\calM$ in \eqref{eq: convex set}, it can be seen that $\mM^\natural\widehat{\mQ}_0 \in\calM$, where $\widehat{\mQ}_0 = \arg\min\limits_{\mQ} \fronorm{\hM_0 - \mM^\natural\mQ}$.  
Thus we have
\begin{align}
	\label{reduce projection}
	\dist(\mM_0, \mM^\natural) =&\min_{\mQ}\fronorm{\mM_0 - \mM^\natural\mQ} \notag\\
	\leq &\fronorm{\mM_0 - \mM^\natural\widehat{\mQ}_0} \notag\\
	=&  \fronorm{\calP_{\calM} \left( \hM_0 - \mM^\natural\widehat{\mQ}_0 \right)}\notag\\
	\leq &\fronorm{\hM_0 - \mM^\natural\widehat{\mQ}_0 } \notag\\
	=&\dist(\hM_0, \mM^\natural).
\end{align}
To complete the proof, it suffices to control $\dist(\hM_0, \mM^\natural)$. A straightforward computation yields that
\begin{align*}
	\dist^2(\hM_0, \mM^\natural)\stackrel{(a)}{\leq } &\frac{1}{2(\sqrt{2} -1  )\sigma_r^2(\mM^\natural)} \fronorm{ \hM_0\hM_0^\tranH - \mM^\natural {\mM^\natural}^\tranH}^2\\
	\stackrel{(b)}{=}&\frac{1}{4(\sqrt{2} -1)\sigma_r} \fronorm{ \hM_0\hM_0^\tranH - \mM^\natural {\mM^\natural}^\tranH}^2\\	
	\stackrel{(c)}{\leq } &\frac{1}{4(\sqrt{2} -1  )\sigma_r}  \cdot 4\fronorm{\hZ_0 - \mZ^\natural}^2\\
	\leq  &\frac{1}{(\sqrt{2} -1  )\sigma_r}  \cdot 2 r\opnorm{\hZ_0 - \mZ^\natural}^2\\
	\leq &\frac{2r}{ (\sqrt{2} -1  )\sigma_r } \cdot   \frac{\mu_0\mu sr\log^2(sn)}{n} \sigma_1^2\\
	\leq& \frac{\varepsilon^2\sigma_r}{\mu_0s} 
\end{align*} 
provided that $n\geq c_0\varepsilon^{-2}\kappa^2 \mu_0^2 \mu s^2 r^2 \log^2(sn)$, where  step (a) is due to Lemma 5.4 in \cite{tu2016low}, step (b) has used the fact $\sigma_r(\mM^\natural) = \sqrt{2}\sigma_r$, and the step (c) {\color{black}{can be derived as follows.

Let $\mA_1, \mB_1\in \C^{sn_1\times r}$ and $\mA_2, \mB_2\in\C^{n_2\times r}$ be four complex matrices. It is direct to obtain that
\begin{align*}
    \la\mA_1\mA_1^\tranH, \mB_1\mB_1^\tranH\ra + \la\mA_2\mA_2^\tranH, \mB_2\mB_2^\tranH\ra=&\la\mA_1^\tranH\mB_1, \mA_1^\tranH\mB_1\ra + \la\mA_2^\tranH\mB_2, \mA_2^\tranH\mB_2\ra\\
    \geq &2\Re \lb\la\mA_1\mA_2^\tranH, \mB_1\mB_2^\tranH\ra\rb.\numberthis\label{ineq: inner}
\end{align*}
Then a simple calculation yields that
\begin{align*}
    \fronorm{ \hM_0\hM_0^\tranH - \mM^\natural {\mM^\natural}^\tranH}^2 = &2\fronorm{\hU_0\hbSigma_0\hV_0^\tranH - \mU\bSigma\mV^\tranH}^2 + \fronorm{\hU_0\hbSigma_0\hU_0^\tranH - \mU\bSigma\mU^\tranH}^2 + \fronorm{\hV_0\hbSigma_0\hV_0^\tranH - \mV\bSigma\mV^\tranH}^2.
\end{align*}
Denote $\mA_1 = \hU_0{\hbSigma_0}^{1/2}$, $\mB_1 = \mU{\bSigma}^{1/2}$, $\mA_2 = \hV_0{\hbSigma_0}^{1/2}$ and $\mB_2 = \mV{\bSigma}^{1/2}$. Applying \eqref{ineq: inner} can be easily verified that
\begin{align*}
    \fronorm{\hU_0\hbSigma_0\hU_0^\tranH - \mU\bSigma\mU^\tranH}^2 + \fronorm{\hV_0\hbSigma_0\hV_0^\tranH - \mV\bSigma\mV^\tranH}^2 \leq &2\fronorm{\hU_0\hbSigma_0\hV_0^\tranH - \mU\bSigma\mV^\tranH}^2.
\end{align*}
Hence 
\begin{align*}
    \fronorm{ \hM_0\hM_0^\tranH - \mM^\natural {\mM^\natural}^\tranH}^2 &\leq 4\fronorm{\hU_0\hbSigma_0\hV_0^\tranH - \mU\bSigma\mV^\tranH}^2\\
    & = 4\fronorm{\hZ_0 - \mZ^\natural}^2.
\end{align*}

}

} Thus we complete the proof of Lemma \ref{lemma: initalization error}.


\subsection{Proof of Lemma \ref{lem: local convergence}}
\label{proof local convergence}
We first establish a regularity condition in the following lemma whose proof is provided in Section \ref{proof key lemma}. 
\begin{lemma}
	\label{key lemma}
	Let $\bDelta = \mM - \mM^\natural \mQ$, where $\mQ = \arg\min\limits_{\mQ\mQ^\tranH = \mQ^\tranH\mQ = \mI_r}\fronorm{\mM- \mM^\natural \mQ}$. Then one has 
	\begin{align}
		\label{ineq: correlation condition}
		\Re \lb\la \nabla f(\mM), \bDelta \ra\rb \geq \frac{\eta}{2} \fronorm{\nabla f(\mM) }^2+ \frac{\beta}{2} \fronorm{\bDelta}^2  
	\end{align}
	happens with high probability for all $\mM$ such that $\fronorm{\bDelta}^2\leq \frac{\varepsilon^2\sigma_r}{\mu_0 s}$, where $\eta\leq \frac{\sigma_r}{9000(\mu_0 \mu sr\sigma_1)^2}$, and $\beta = \frac{\sigma_r}{72}$.
\end{lemma}

Let $\hM_{t+1} = \mM_t - \eta \nabla f(\mM_t)$. Under the condition of \eqref{ineq: correlation condition}, one has 
\begin{align*}
	\dist^2(\mM_{t+1}, \mM^\natural) \stackrel{(a)}{\leq } &\dist^2(\hM_{t+1},\mM^\natural)\\
	\leq& \fronorm{\hM_{t+1}-\mM^\natural\mQ_{t}}^2\\
	=& \fronorm{\bDelta_t - \eta \nabla f(\mM_t)}^2\\
	=& \fronorm{\bDelta_t}^2 + \eta^2 \fronorm{\nabla f(\mM_t)}^2 - 2\eta\Re \lb\la \bDelta_t , \nabla f(\mM_t) \ra\rb \\
	\stackrel{(b)}{\leq} &\fronorm{\bDelta_t}^2+ \eta^2 \fronorm{\nabla f(\mM_t)}^2  -\eta^2 \fronorm{\nabla f(\mM_t) }^2 - \beta \eta \fronorm{\bDelta_t}^2  \\
	=&(1-\eta\beta) \fronorm{\bDelta_t}^2  ,
\end{align*}
where step (a) follows the same argument as \eqref{reduce projection} and step (b) is due to \eqref{ineq: correlation condition}. Then a little algebra yields that
\begin{align*}
	\dist^2(\mM_{t+1}, \mM^\natural) &\leq (1-\eta\beta)^{t+1} \dist^2(\mM_0,\mM^\natural)\\
	&\leq (1-\eta\beta)^{t+1} \frac{\varepsilon^2\sigma_r}{\mu_0 s},
\end{align*}
which completes the proof of Lemma \ref{lem: local convergence}.



\subsection{Proof of Lemma \ref{lem: init opnorm}}
\label{proof lemma init opnorm}
Notice that $\calG\calD\calAT(\vy) = \calG\calAT\mD(\vy)=  \calG\calAT\calA\calGT(\mZ^\natural) $. A simple computation yields that
\begin{align*}
	\E{\calG\calAT\calA\calGT(\mZ^\natural)} &= \E{\calG\lb\sum_{i=0}^{n-1}\la\vb_i\ve_i^\tran,\calGT(\mZ^\natural)\ra\vb_i\ve_i^\tran\rb}\\
	&= \calG\lb\sum_{i=0}^{n-1}\E{\vb_i\vb_i^\tranH\calGT(\mZ^\natural)\ve_i\ve_i^\tran}\rb\\
	&= \calG\calGT(\mZ^\natural) = \mZ^\natural,
\end{align*}
where the third equality is due to the isotropy property of $\{\vb_i\}_{i=0}^{n-1}$. Let us first bound
\begin{align*}\opnorm{\calG\calAT\calA\calGT(\mZ^\natural) - \E{\calG\calAT\calA\calGT(\mZ^\natural)}}
\end{align*}
by the matrix Bernstein inequality \eqref{ineq: bernstein}. The matrix $\calG\calAT\calA\calGT(\mZ^\natural) - \E{\calG\calAT\calA\calGT(\mZ^\natural)}$ can be rewritten as 
\begin{align*}
	\calG\calAT\calA\calGT(\mZ^\natural) - \E{\calG\calAT\calA\calGT(\mZ^\natural)} =& \sum_{i=0}^{n-1} \calG\lb(\vb_i\vb_i^\tranH - \mI_s)\calGT(\mZ^\natural)\ve_i\ve_i^\tran\rb\\
	= &\sum_{i=0}^{n-1} \mG_i\otimes\lb(\vb_i\vb_i^\tranH-\mI_s)\calGT(\mZ^\natural)\ve_i\rb\\
	=&: \sum_{i=0}^{n-1} \mY_i.
\end{align*}
Notice that $\{\mY_i\}_{i=0}^{n-1}$ are independent mean-zero random matrices with 
\begin{align*}
	\opnorm{\mY_i} &= \opnorm{\mG_i\otimes\lb(\vb_i\vb_i^\tranH-\mI_s)\calGT(\mZ^\natural)\ve_i\rb}\\
	&\leq \opnorm{\mG_i}\cdot\opnorm{(\vb_i\vb_i^\tranH - \mI_s)\calGT(\mZ^\natural)\ve_i}\\
	&\leq \frac{1}{\sqrt{\omega_i}}\cdot \max\{\vecnorm{\vb_i}{2}^2,1\}\cdot \vecnorm{\calGT(\mZ^\natural)\ve_i}{2}\\
	&\stackrel{(a)}{\leq} s\mu_0\cdot \max_i \frac{1}{\sqrt{\omega_i}} \vecnorm{\calGT(\mZ^\natural)\ve_i}{2}\\
	&\stackrel{(b)}{\leq }s\mu_0\cdot  \frac{\mu_1 r}{n}\sigma_1,
\end{align*}
where step (a) follows from \eqref{eq:incoherence} and step (b)  is due to Lemma~\ref{lem: two norms}. Moreover, letting $\vw_i := (\vb_i\vb_i^\tranH-\mI_s)\calGT(\mZ^\natural)\ve_i$, we have 
\begin{align*}
	\opnorm{\E{\sum_{i=0}^{n-1}\mY_i^\tranH\mY_i}} &= \opnorm{\sum_{i=0}^{n-1}\E{(\mG_i\otimes\vw_i)^\tranH(\mG_i\otimes\vw_i)}}\\
	&=\opnorm{\sum_{i=0}^{n-1}(\mG_i^\tranH\mG_i)\cdot \E{\vecnorm{\vw_i}{2}^2}}\\
	&\leq \sum_{i=0}^{n-1} \opnorm{\mG_i^\tranH\mG_i}\cdot  \E{\vecnorm{\vw_i}{2}^2} \\
	&\stackrel{(a)}{\leq}\sum_{i=0}^{n-1} \frac{1}{w_i} \cdot s\mu_0  \vecnorm{\calGT(\mZ^\natural)\ve_i}{2}^2\\
	&\stackrel{(b)}{\leq } s\mu_0\cdot \sqrt{\frac{\mu_1 r\log(sn)}{n}}\sigma_1,
\end{align*}
where step (a) follows from $\opnorm{\mG_i}\leq 1/\sqrt{w_i}$ and the fact $\E{\vecnorm{\vw_i}{2}^2} \leq s\mu_0\cdot \vecnorm{\calGT(\mZ^\natural)\ve_i}{2}^2$, and step (b) is due to Lemma \ref{lem: two norms}. Moreover, the fact used in step (a) can be proved as follows:
\begin{align*}
	\E{\vecnorm{\vw_i}{2}^2} &= \E{\ve_i^\tran\lb\calGT(\mZ^\natural)\rb^\tranH(\vb_i\vb_i^\tranH - \mI_s)^2\calGT(\mZ^\natural)\ve_i}\\
	&= \ve_i^\tran(\calGT(\mZ^\natural))^\tranH\E{(\vb_i\vb_i^\tranH - \mI_s)^2}\calGT(\mZ^\natural)\ve_i\\
	&= \ve_i^\tran(\calGT(\mZ^\natural))^\tranH\E{ \vecnorm{\vb_i}{2}^2\vb_i\vb_i^\tranH - \mI_s}\calGT(\mZ^\natural)\ve_i\\
	&\leq (\mu_0 s - 1) \cdot \vecnorm{\calGT(\mZ^\natural)\ve_i}{2}^2\\
	&\leq \mu_0 s\vecnorm{\calGT(\mZ^\natural)\ve_i}{2}^2,
\end{align*}
where the third line is due to \eqref{eq: isotropy property}. Similarly, one has
\begin{align*}
	\opnorm{\E{\sum_{i=0}^{n-1}\mY_i\mY_i^\tranH}} &\leq \sum_{i=0}^{n-1} \opnorm{\mG_i}^2\cdot \opnorm{\E{\vw_i\vw_i^\tranH}}\\
	&\leq \sum_{i=0}^{n-1} \opnorm{\mG_i}^2\cdot \E{ \twonorm{\vw_i}^2 }\\
	&\leq s\mu_0\cdot \sqrt{\frac{\mu_1 r\log(sn)}{n}}\sigma_1.
\end{align*}
Applying the matrix Bernstein inequality \eqref{ineq: bernstein}  shows that, with probability greater than $1-(sn)^{-c_1}$,  
\begin{align*}
	\opnorm{\calG\calAT\calA\calGT(\mZ^\natural) - \E{\calG\calAT\calA\calGT(\mZ^\natural)}}
\leq &c_1\sqrt{s\mu_0\log(sn)}\cdot\sqrt{\frac{\mu_1 r\log(sn)}{n}}\sigma_1+ c_2 s\mu_0\log(sn)\cdot \frac{\mu_1 r}{n}\sigma_1\\
\leq& c_3\sqrt{\frac{\mu_0\mu_1 sr\log^2(sn)}{n}}\sigma_1
\end{align*}
provided $n\geq c_0 \mu_0\mu sr\log^2(sn)$. Therefore, the event
\begin{align*}
	\opnorm{\hZ_0 - \mZ^\natural} =& \opnorm{\calP_r\left(\calG\calA\calAT\calGT(\mZ^\natural)  \right)-\mZ^\natural}\\
	\leq& \opnorm{\calP_r\left(\calG\calA\calAT\calGT(\mZ^\natural) \right)- \calG\calAT\calA\calGT(\mZ^\natural)}\\
	&\quad  + \opnorm{\calG\calAT\calA\calGT(\mZ^\natural) - \mZ^\natural}\\
	\leq& ~2\opnorm{\calG\calAT\calA\calGT(\mZ^\natural) - \mZ^\natural}\\
	\leq &~c_0\sqrt{\frac{\mu_0\mu sr\log^2(sn)}{n}}\sigma_1.
\end{align*}
occurs with probability at least $1-(sn)^{-c_1}$. Finally we complete the proof.

\subsection{Proof of Lemma \ref{key lemma}}
\label{proof key lemma}
The proof includes two parts. We will show that
\begin{align}
    \label{key lower bound}
	\Re \lb\la \nabla f(\mM), \bDelta \ra\rb \geq \frac{1}{72}\sigma_r \fronorm{\bDelta}^2 
	+ \frac{1}{8} \fronorm{{\mM^\natural}^\tranH\mS\bDelta }^2
\end{align}
and 
\begin{align}
    \label{key upper bound}
	\fronorm{\nabla f(\mM)}^2 \leq 125(\mu_0\mu s r \sigma_1)^2\fronorm{\bDelta}^2  
	+ \frac{1}{2}\sigma_1\fronorm{{\mM^\natural}^\tranH\mS\bDelta}^2
\end{align}
provided $\varepsilon \leq \frac{1}{3}$. By the assumption 
\begin{align*}
	\eta \leq \frac{\sigma_r}{9000(\mu_0 \mu sr\sigma_1)^2} \leq \frac{1}{2\sigma_1} \text{ and }\beta= \frac{\sigma_r}{72},
\end{align*}
we have
\begin{align*}
	\frac{\eta}{2}\fronorm{\nabla f(\mM)}^2 + \frac{\beta}{2}\fronorm{\bDelta}^2\leq &\lb\frac{1}{2} \frac{\sigma_r}{9000(\mu_0 \mu sr\sigma_1)^2} \cdot 125(\mu_0\mu s r \sigma_1)^2 + \frac{1}{2}\frac{\sigma_r}{72}\rb\fronorm{\bDelta}^2 + \frac{1}{2}\cdot \frac{1}{2\sigma_1}\cdot \frac{1}{2}\sigma_1\fronorm{{\mM^\natural}^\tranH\mS\bDelta}^2 \\
\leq &\frac{\sigma_r}{72}\fronorm{\bDelta}^2 + \frac{1}{8}\fronorm{ {\mM^\natural}^\tranH \mS\bDelta}^2 \\
\leq &\Re \lb\la \nabla f(\mM), \bDelta \ra\rb.
\end{align*}

\subsubsection{Proof of \eqref{key lower bound}}
Let $\nabla f_1, \nabla f_2$ and $\nabla f_3$ be the matrices given by 
\begin{align*}
	\nabla f_1 &= \begin{bmatrix}
		\left(\calG\calAT\left(\calA\calGT(\mL\mR^\tranH) - \mD\vy\right)\right)\mR\\
		\left(\calG\calAT\left(\calA\calGT(\mL\mR^\tranH) - \mD\vy\right)\right)^\tranH\mL\\
	\end{bmatrix}, \\
\nabla f_2 &= \begin{bmatrix}
		\lb(\calI - \calG\calGT)(\mL\mR^\tranH)\rb\mR\\
		\lb(\calI - \calG\calGT)(\mL\mR^\tranH)\rb^\tranH\mL\\
	\end{bmatrix},\\
	\nabla f_3 &= \frac{1}{4}\begin{bmatrix}
		\mL(\mL^\tranH\mL - \mR^\tranH\mR)\\
		\mR(\mR^\tranH\mR - \mL^\tranH\mL)
	\end{bmatrix}.
\end{align*}
A straightforward computation yields that 
\begin{align*}
	\Re \lb\la \nabla  f(\mM), \bDelta \ra\rb &= \Re\lb\la \nabla f_1, \bDelta\ra\rb  +	\Re \lb\la \nabla f_2, \bDelta \ra\rb + \Re\lb\la \nabla f_3, \bDelta \ra\rb.
\end{align*}
We will bound these three terms separately. For the sake of simplification,  let $\bDelta = \begin{bmatrix}
	\bDelta_{\mL}^\tranH &\bDelta_{\mR}^\tranH
\end{bmatrix}^\tranH$ where $\bDelta_{\mL} = \mL - \mL^\natural \mQ \in\C^{sn_1\times r} $ and  $\bDelta_{\mR} = \mR - \mR^\natural \mQ\in\C^{n_2\times r}$.  
\paragraph{Bounding $\Re\lb\la \nabla f_1, \bDelta \ra\rb$}	By applying $\mD\vy = \calA\calGT(\mZ^\natural)  $, we can rewrite $\Re\lb\la \nabla f_1, \bDelta \ra\rb$ as 
	\begin{align*}
		\Re \lb\la \nabla f_1, \bDelta \ra\rb 
		= &\Re\lb\la \calG\calAT\calA\calGT(\mL\mR^\tranH - \mL^\natural {\mR^\natural}^\tranH) , \bDelta_{\mL} \mR^\tranH + \mL\bDelta_{\mR}^\tranH \ra\rb. \numberthis\label{nabla f1}
	\end{align*}
	Notice that 
	\begin{align}
		\label{eq: transform 1}
		\mL\mR^\tranH - \mL^\natural {\mR^\natural}^\tranH \notag
		=&\left( \bDelta_{\mL} +  \mL^\natural \mQ\right)\left( \bDelta_{\mR} +  \mR^\natural \mQ \right)^\tranH -\mL^\natural {\mR^\natural}^\tranH \notag \\
		=& \underbrace{\bDelta_{\mL}\bDelta_{\mR}^\tranH}_{:=\bPsi} + \underbrace{\bDelta_{\mL} {(\mR^\natural \mQ )}^\tranH + (\mL^\natural\mQ) \bDelta_{\mR}^\tranH}_{:=\bPhi}. 
	\end{align}
	and 
	\begin{align}
		\label{eq: transform 2}
		\bDelta_{\mL} \mR^\tranH + \mL\bDelta_{\mR}^\tranH \notag
		=&\bDelta_{\mL} (\bDelta_{\mR} + \mR^\natural \mQ)^\tranH + (\bDelta_{\mL } + \mL^\natural \mQ)\bDelta_{\mR}^\tranH \notag \\
		=&2 \bDelta_{\mL} \bDelta_{\mR}^\tranH +  \bDelta_{\mL}(\mRQ)^\tranH + (\mLQ)\bDelta_{\mR}^\tranH \notag\\
		=&2\bPsi + \bPhi. 
	\end{align}
	Then $\Re \lb\la \nabla f_1, \bDelta \ra\rb$ can be bounded as follows:
	\begin{align*}
		\Re \lb\la \nabla f_1, \bDelta \ra\rb
		=&\Re\lb\la \calA\calGT\left( \bPhi + \bPsi \right), \calA\calGT\left( \bPhi + 2 \bPsi \right)\ra\rb \\
		=& \twonorm{\calA\calGT\left( \bPhi   \right)}^2 + 2\twonorm{ \calA\calGT\left(  \bPsi \right)}^2  + 3 \Re\lb\la\calA\calGT\left( \bPhi  \right), \calA\calGT\left( \bPsi \right) \ra\rb  \\
		\geq& \twonorm{\calA\calGT\left( \bPhi   \right)}^2 + 2\twonorm{ \calA\calGT\left(  \bPsi \right)}^2 - 3 \left|\la\calA\calGT\left( \bPhi  \right), \calA\calGT\left( \bPsi \right) \ra \right| \\
		\geq &\twonorm{\calA\calGT\left( \bPhi   \right)}^2 + 2\twonorm{ \calA\calGT\left(  \bPsi \right)}^2- 3\twonorm{\calA\calGT\left( \bPhi  \right)} \cdot \twonorm{ \calA\calGT\left( \bPsi \right) }\\
		\geq&~ \frac{3}{4}\twonorm{ \calA\calGT\left( \bPhi   \right) }^2 - 7\twonorm{ \calA\calGT\left(  \bPsi \right)}^2\\
		\geq&~ \frac{3}{4}\twonorm{ \calA\calGT\left( \bPhi   \right) }^2 - \frac{7\varepsilon^2 }{4}\sigma_r \fronorm{\bDelta}^2, \numberthis\label{T1}
	\end{align*}
	where the second line is due to $\la \vx,\vx\ra = \twonorm{\vx}^2$ and $\Re\lb\la \vx, \vy\ra\rb = \Re\lb\la \vy, \vx\ra\rb$ for any vectors $\vx,\vy\in\C^n$, the third line is due to $\Re\lb\la \vx, \vy\ra\rb \leq |\la \vx ,\vy\ra|$, and the last second line is due to $a+2b-3\sqrt{ab} \geq \frac{3}{4}a - 7b$ for any $a,b\geq 0$ and the last line is due to the fact $\twonorm{ \calA\calGT\left(  \bPsi \right)}^2  \leq \frac{\varepsilon^2 }{4}\sigma_r \fronorm{\bDelta}^2$. Moreover the fact used in the last line can be proved as follows:
	\begin{align*}
		\twonorm{ \calA\calGT\left(  \bPsi \right)}^2 &=\twonorm{\calA\calGT\left( \bDelta_{\mL}\bDelta_{\mR}^\tranH  \right)}^2\\
		&\leq \left( \opnorm{\calA}\opnorm{\calGT}\cdot \fronorm{\bDelta_{\mL}\bDelta_{\mR}^\tranH  } \right)^2\\
		&\leq \mu_0 s \cdot \frac{1}{4} \left( \fronorm{\bDelta_{\mL}}^2 + \fronorm{ \bDelta_{\mR}}^2\right)^2\\
		&=\frac{\mu_0 s}{4}\fronorm{\bDelta}^4\\
		&\leq \frac{\varepsilon^2 }{4}\sigma_r \fronorm{\bDelta}^2, \numberthis \label{upper bound a1}
	\end{align*}
	where the second line is due to $\opnorm{\calA} \leq \sqrt{\mu_0 s}$, $\opnorm{\calGT}\leq 1$ and the last line is due to $\fronorm{\bDelta}^2 \leq \frac{\varepsilon^2\sigma_r}{\mu_0s}$.  Plugging \eqref{T1} into \eqref{nabla f1} reveals that 
	\begin{align}
		\label{ineq: lower bound of Re gls}
		\Re \lb\la \nabla f_1, \bDelta \ra\rb  \geq \frac{3}{4}\twonorm{\calA\calGT\left( \bPhi   \right)}^2  - \frac{7\varepsilon^2}{4}\sigma_r \fronorm{\bDelta}^2.
	\end{align}
	
\paragraph{Bounding $\Re\lb\la \nabla f_2, \bDelta\ra\rb$}
	This term can be bounded as follows:
	\begin{align*}
		\Re\lb\la \nabla f_2, \bDelta\ra\rb
		=&\Re \lb\la (\calI - \calG\calGT)(\mL\mR^\tranH), \bDelta_{\mL} \mR^\tranH + \mL\bDelta_{\mR}^\tranH\ra\rb\\
		= &\Re \lb\la (\calI - \calG\calGT)(\mL\mR^\tranH - \mL^\natural {\mR^\natural}^\tranH), \bDelta_{\mL} \mR^\tranH + \mL\bDelta_{\mR}^\tranH\ra\rb\\ 
		=&\Re \lb\la (\calI - \calG\calGT)(\bPhi + \bPsi), (\bPhi + 2\bPsi)\ra\rb\\ 
		\geq &\fronorm{\left( \calI - \calG\calGT \right) \left( \bPhi\right) }^2 + 2\fronorm{\left( \calI - \calG\calGT \right) (\bPsi)}^2 \\
		&\quad -3\fronorm{\left( \calI - \calG\calGT \right) \left( \bPhi\right) } \cdot \fronorm{\left( \calI - \calG\calGT \right) (\bPsi)}\\
		\geq& \frac{3}{4}\fronorm{\left( \calI - \calG\calGT\right) \left( \bPhi\right) }^2 - 7\fronorm{\left( \calI - \calG\calGT\right) \left(\bPsi\right) }^2,\numberthis\label{ineq: lower bound if gh}
	\end{align*}
	where the second line is due to $(\calI - \calG\calGT)(\mL^\natural {\mR^\natural}^\tranH)=\bzero$ and the last line follows from the fact that $a+2b-3\sqrt{ab} \geq \frac{3}{4}a - 7b$ for any $a,b\geq 0$. Combining \eqref{ineq: lower bound of Re gls} and  \eqref{ineq: lower bound if gh} together, one has
	\begin{align*}
		\Re \lb\la \nabla f_1, \bDelta \ra\rb + \Re\lb\la \nabla f_2, \bDelta \ra\rb
		\geq&  \frac{3}{4}\twonorm{\calA\calGT\left( \bPhi   \right)}^2  -  \frac{7\varepsilon^2}{4}\sigma_r \fronorm{\bDelta}^2 + \frac{3}{4}\fronorm{\left( \calI - \calG\calGT\right) \left( \bPhi\right) }^2 - 7\fronorm{\left( \calI - \calG\calGT \right) (\bPsi) }^2  \\
		\geq& \frac{3}{4} \left( \twonorm{\calA\calGT\left( \bPhi   \right)}^2  + \fronorm{\left( \calI - \calG\calGT \right) \left( \bPhi\right) }^2  \right) -  \frac{7\varepsilon^2}{4}\sigma_r \fronorm{\bDelta}^2 - 7\fronorm{  \bPsi  }^2 ,\numberthis\label{ineq: lower bound of ls and h}
	\end{align*}
	where the last line follows from the fact that $\calI - \calG\calGT $ is a projection operator. Let $T$ be the tangent space at $\mZ^\natural$ defined as follows
	\begin{align*}
		T := \{\mU\mJ^\tranH + \mK\mV^\tranH: \mJ\in\C^{n_2\times r}, \mK\in\C^{sn_1\times r}\}.
	\end{align*} 
	It can be seen that $\bPhi\in T$. Therefore  a simple calculation yields that 
	\begin{align*}
		\twonorm{\calA\calGT\left( \bPhi   \right)}^2 + \fronorm{\left( \calI - \calG\calGT\right) \left( \bPhi\right) }^2 
		=& \la \calG\calAT\calA\calGT(\bPhi), \bPhi \ra + \la \bPhi, \bPhi \ra - \la \calG\calGT(\bPhi), \bPhi \ra \\
		\geq& \fronorm{\bPhi}^2 - \left| \la  \calG\left( \calI - \calAT\calA \right) \calGT(\bPhi), \bPhi  \ra\right|\\
		= &\fronorm{\bPhi}^2 - \left| \la  \calP_T\calG\left( \calI - \calAT\calA \right) \calGT\calP_T(\bPhi), \bPhi  \ra\right|\\
		\geq& (1-\varepsilon)  \fronorm{\bPhi}^2,\numberthis\label{ineq: fronorm of bPhi}
	\end{align*}
	where the last line follows from Lemma~\ref{lem: pt opnorm} and the assumption on $n$. In addition, we have
	\begin{align*}
		\fronorm{\bPsi}^2 \leq&  \left( \fronorm{  \bDelta_{\mL} } \fronorm{\bDelta_{\mR}} \right)^2 \\
		\leq &\frac{1}{4} \fronorm{\bDelta}^4\\
		\leq &\frac{\varepsilon^2 \sigma_r}{4} \fronorm{\bDelta}^2,\numberthis\label{ineq: fronorm of bPsi}
	\end{align*}
	where we use $\fronorm{\bDelta}^2\leq \frac{\varepsilon^2 \sigma_r(\mZ^\natural)}{s\mu_0}\leq \varepsilon^2 \sigma_r(\mZ^\natural)$.
	Hence plugging \eqref{ineq: fronorm of bPhi} and \eqref{ineq: fronorm of bPsi} into \eqref{ineq: lower bound of ls and h} yields that 
	\begin{align*}
		\Re \lb\la \nabla f_1, \bDelta \ra\rb + \Re\lb\la \nabla f_2, \bDelta \ra\rb  
		\geq&  \frac{3}{4} (1-\varepsilon)  \fronorm{\bPhi}^2 - \frac{7\varepsilon^2}{2}\sigma_r \fronorm{\bDelta}^2  \\
		\geq& \frac{1}{2}\fronorm{\bPhi}^2 - \frac{7\varepsilon^2}{2}\sigma_r \fronorm{\bDelta}^2 \\
		=& \frac{1}{2} \lb \fronorm{\bDelta_{\mL} (\mRQ)^\tranH}^2 + \fronorm{\mLQ\bDelta_{\mR}^\tranH }^2 \rb -  \frac{7\varepsilon^2}{2}\sigma_r \fronorm{\bDelta}^2\\
		& \quad + \Re\lb\la  \bDelta_{\mL} (\mRQ)^\tranH, (\mLQ)\bDelta_{\mR}^\tranH\ra\rb  \\
		\geq &\lb \frac{1}{2}-\frac{7\varepsilon^2}{2} \rb \sigma_r \fronorm{\bDelta}^2   + \Re\lb\la \bDelta_{\mL}^\tranH (\mLQ), (\mRQ)^\tranH \bDelta_{\mR} \ra\rb \\
		\geq& \frac{1}{9} \sigma_r \fronorm{\bDelta}^2 + \Re\lb\la \bDelta_{\mL}^\tranH(\mLQ), (\mRQ)^\tranH \bDelta_{\mR} \ra\rb,\numberthis\label{ineq: two terms}
	\end{align*}
	where the last line is due to $\varepsilon\leq \frac{1}{3}$.
	\paragraph{Bounding $\Re\lb\la \nabla f_3, \bDelta\ra\rb $}
	Denote $$\mS=\begin{bmatrix}
		 \mI_{sn_1} &\\
		 &-\mI_{n_2}\\
	\end{bmatrix}.$$
	{\color{black}{We can bound $\Re\lb\la \nabla f_3, \bDelta\ra\rb $ as follows:
	\begin{align*}
		4\Re\lb\la \nabla f_3, \bDelta \ra\rb
		 =&  \Re \lb\la \mS\mM\mM^\tranH\mS\mM, \bDelta\ra\rb \\
		=&\Re\lb\la \mM^\tranH\mS\mM,\mM^\tranH\mS\bDelta\ra\rb \\
		=& \Re \lb\la (\bDelta + \mM^\natural\mQ)^\tranH\mS(\bDelta + \mM^\natural\mQ),(\bDelta + \mM^\natural\mQ)^\tranH\mS\bDelta\ra\rb \\
		=& \fronorm{\bDelta^\tranH\mS\bDelta}^2 + \fronorm{(\mM^\natural\mQ)^\tranH\mS\bDelta }^2+ 3\Re\lb\la(\mM^\natural\mQ)^\tranH\mS\bDelta,\bDelta^\tranH\mS\bDelta\ra\rb+ \Re\lb\la(\mM^\natural\mQ)^\tranH\mS\bDelta,\bDelta^\tranH\mS(\mM^\natural\mQ)\ra\rb \\
		=& \frac{1}{2}\fronorm{(\mM^\natural\mQ)^\tranH\mS\bDelta }^2 + \frac{1}{2}\fronorm{(\mM^\natural\mQ)^\tranH\mS\bDelta+3\bDelta^\tranH\mS\bDelta}^2- \frac{7}{2} \fronorm{\bDelta^\tranH\mS\bDelta}^2 + \Re\la(\mM^\natural\mQ)^\tranH\mS\bDelta,\bDelta^\tranH\mS(\mM^\natural\mQ)\ra\\
		=& \frac{1}{2}\fronorm{(\mM^\natural\mQ)^\tranH\mS\bDelta }^2 + \frac{1}{2}\fronorm{(\mM^\natural\mQ)^\tranH\mS\bDelta+3\bDelta^\tranH\mS\bDelta}^2- \frac{7}{2} \fronorm{\bDelta^\tranH\mS\bDelta}^2 \\
		&\quad + \Re\lb\la(\mM^\natural\mQ)^\tranH\bDelta,\bDelta^\tranH(\mM^\natural\mQ)\ra\rb - 4\Re\lb\la \bDelta_{\mL}^\tranH(\mLQ), (\mRQ)^\tranH\bDelta_{\mR} \ra\rb\\
		\geq& \frac{1}{2} \fronorm{(\mM^\natural\mQ)^\tranH\mS\bDelta }^2 - \frac{7}{2}\fronorm{\bDelta^\tranH\mS\bDelta}^2 -4\Re\lb\la \bDelta_{\mL}^\tranH(\mLQ), (\mRQ)^\tranH\bDelta_{\mR} \ra\rb. \numberthis \label{ineq: lower bound of gb}
	\end{align*} 
	The fourth equality is due to $(\mM^\natural\mQ)^\tranH\mS(\mM^\natural\mQ) = \bzero$. The last equality follows from 
	\begin{align*}
	    \Re\lb\la \bDelta_{\mL}^\tranH(\mLQ), (\mRQ)^\tranH\bDelta_{\mR} \ra\rb =& \Re\lb\la (\mRQ)^\tranH\bDelta_{\mR}, \bDelta_{\mL}^\tranH(\mLQ) \ra\rb\\
=& \Re\lb\la (\mLQ)^\tranH\bDelta_{\mL}, \bDelta_{\mR}^\tranH(\mRQ) \ra\rb.
	\end{align*}
	}}
\\
Finally, combining \eqref{ineq: lower bound of gb} with \eqref{ineq: two terms} yields that 
\begin{align*}
	\Re \lb\la \nabla f(\mM), \bDelta \ra\rb 
	\geq &\frac{1}{9} \sigma_r \fronorm{\bDelta}^2 + \frac{1}{8} \fronorm{{\mM^\natural}^\tranH\mS\bDelta }^2- \frac{7}{8}\fronorm{\bDelta^\tranH\mS\bDelta}^2 \\
	\geq &\frac{1}{72}\sigma_r \fronorm{\bDelta}^2 + \frac{1}{8} \fronorm{{\mM^\natural}^\tranH\mS\bDelta }^2,\numberthis\label{ineq: lower bound of Re}
\end{align*}
where the last line follows from that $\fronorm{\bDelta^\tranH\mS\bDelta}^2 \leq \fronorm{\bDelta}^4 \leq \frac{\varepsilon^2 \sigma_r}{\mu_0 s}\fronorm{\bDelta}^2$ and $\varepsilon\leq \frac{1}{3}$.

\subsubsection{Proof of \eqref{key upper bound}}
Applying simple triangular inequality yields that 
\begin{align*}
	\fronorm{\nabla f(\mM)}^2&=\fronorm{\nabla f_1 + \nabla f_2+ \nabla f_3}^2\\
	&\leq 2\fronorm{\nabla f_1 + \nabla f_2}^2 + 2\fronorm{  \nabla f_3}^2\\
	&\leq 4\fronorm{\nabla f_1 }^2 + 4\fronorm{\nabla f_2 }^2 + 2 \fronorm{  \nabla f_3}^2.\numberthis\label{ineq: upper bound of gradient of fronorm of g}
\end{align*}
We will bound these three terms separately.
\paragraph{Bounding $\fronorm{\nabla f_1 }^2$}
	For any $$\mX = \begin{bmatrix} \mX_{\mL}^\tranH & \mX_{\mR}^\tranH \end{bmatrix}^\tranH\in\C^{(s n_1+n_2)\times r}$$ such that $\fronorm{\mX}=1$, we have
	\begin{align*}
		\lab\la\nabla f_1, \mX\ra\rab^2 
		= &\lab \la \calG\calAT\left(\calA\calGT(\mL\mR^\tranH) - \mD\vy\right), \mX_{\mL}\mR^\tranH + \mL\mX_{\mR}^\tranH\ra\rab^2\\
		=&\lab\la\calG\calAT\calA\calGT(\mL\mR^\tranH - \mL^\natural{\mR^\natural}^\tranH),\mX_L\mR^\tranH + \mL\mX_R^\tranH\ra\rab^2\\
		\leq& \big( \lab\la\calG\calAT\calA\calGT(\mL\bDelta_{\mR}^\tranH),\mX_{\mL}\mR^\tranH\ra\rab \\
		&\quad + \lab\la\calG\calAT\calA\calGT(\bDelta_{\mL}(\mRQ)^\tranH),\mX_{\mL}\mR^\tranH\ra\rab\\
		&\quad + \lab\la\calG\calAT\calA\calGT(\mL\bDelta_{\mR}^\tranH),\mL\mX_{\mL}^\tranH\ra\rab \\
		&\quad + \lab\la\calG\calAT\calA\calGT(\bDelta_{\mL}(\mRQ)^\tranH),\mL\mX_{\mR}^\tranH\ra\rab \big)^2.\numberthis\label{ineq: sqrt T3}
	\end{align*}
	where we have used the fact $\mL\mR^\tranH - \mL^\natural {\mR^\natural}^\tranH = \mL\bDelta_{\mR}^\tranH + \bDelta_{\mL}(\mRQ)^\tranH$. 
	Apparently, the upper bounds of the above four terms can be established similarly. We focus on the first term $\lab\la\calG\calAT\calA\calGT(\mL\bDelta_R^\tranH),\mX_L\mR^\tranH\ra\rab$ and its upper bound can be obtained as follows:
	\begin{align*}
		\lab\la\calG\calAT\calA\calGT(\mL\bDelta_R^\tranH),\mX_L\mR^\tranH\ra\rab 
		=& \lab\la \calA\calGT(\mL\bDelta_R^\tranH),\calA\calGT(\mX_L\mR^\tranH)\ra\rab \\
		\leq& \sum_{i=0}^{n-1}  \lab\la\vb_i\ve_i^\tran, \calGT(\mL\bDelta_R^\tranH)\ra\rab\cdot\lab\la\vb_i\ve_i^\tran,\calGT(\mX_L\mR^\tranH)\ra\rab\\
		=&\sum_{i=0}^{n-1}  \lab\la \calG( \vb_i\ve_i^\tran), \mL\bDelta_R^\tranH\ra\rab\cdot\lab\la \calG(\vb_i\ve_i^\tran), \mX_L\mR^\tranH \ra\rab\\
		=&\sum_{i=0}^{n-1}  \lab\la \mG_i \otimes\vb_i , \mL\bDelta_R^\tranH\ra\rab\cdot\lab\la \mG_i \otimes\vb_i ,  \mX_L\mR^\tranH \ra\rab.\numberthis\label{upper bound c}
	\end{align*}
	To complete the proof, it suffers to control $\lab\la \mG_i \otimes\vb_i , \mL\bDelta_R^\tranH\ra\rab$ and $\lab\la \mG_i \otimes\vb_i ,  \mX_L\mR^\tranH \ra\rab$. Notice that
	\begin{align*}
		\lab\la \mG_i \otimes\vb_i , \mL\bDelta_R^\tranH\ra\rab
		=& \left| \frac{1}{\sqrt{w_i}}\sum_{j+k=i} \la (\ve_j\otimes \vb_i)  \ve_k^\tran, \mL\bDelta_R^\tranH\ra \right| \\
		\leq&  \frac{1}{\sqrt{w_i}}\sum_{j+k=i} \twonorm{ (\ve_j^\tran \otimes \vb_i^\tranH)\mL} \twonorm{ \ve_k^\tran \bDelta_R}\\
		\leq& \sqrt{\sum_{j+k=i} \twonorm{ (\ve_j^\tran \otimes \vb_i^\tranH)\mL}^2\twonorm{ \ve_k^\tran \bDelta_R}^2}\\
		= &\sqrt{\sum_{j+k=i} \twonorm{  \vb_i^\tranH \mL_j}^2\twonorm{ \ve_k^\tran \bDelta_R}^2}\\
		\leq& \sqrt{\sum_{j+k=i} \twonorm{  \vb_i}^2\fronorm{ \mL_j}^2\twonorm{ \ve_k^\tran \bDelta_R}^2}\\
		 \leq &\sqrt{\mu_0 s}\cdot \sqrt{\sum_{j+k=i} \fronorm{ \mL_j}^2\twonorm{ \ve_k^\tran \bDelta_R}^2}\\
\leq& \sqrt{\mu_0 s}\cdot \sqrt{\frac{\mu r\sigma }{n}} \cdot \sqrt{\sum_{j+k=i} \twonorm{ \ve_k^\tran \bDelta_R}^2}. \numberthis\label{upper bound a}
	\end{align*}
	Similarly one has
	\begin{align*}
		\lab\la \mG_i \otimes\vb_i ,  \mX_L\mR^\tranH \ra\rab 
		\leq& \sqrt{\mu_0 s}\cdot \sqrt{\sum_{j+k=i} \fronorm{ [\mX_{\mL}]_j}^2\twonorm{ \ve_k^\tran\mR}^2}\\
		\leq& \sqrt{\mu_0 s}\cdot \sqrt{\frac{\mu r\sigma }{n}} \cdot \sqrt{\sum_{j+k=i} \fronorm{ [\mX_{\mL}]_j}^2}.\numberthis\label{upper bound b}
	\end{align*}
	Plugging \eqref{upper bound a} and \eqref{upper bound b} into \eqref{upper bound c} yields that
	\begin{align*}
		\lab\la\calG\calAT\calA\calGT(\mL\bDelta_R^\tranH),\mX_L\mR^\tranH\ra\rab  
		\leq& \frac{\mu_0 s\cdot \mu  r\sigma}{n}\sum_{i=0}^{n-1}   \sqrt{\sum_{j+k=i} \twonorm{ \ve_k^\tran \bDelta_R}^2}\cdot \sqrt{\sum_{j+k=i} \fronorm{ [\mX_{\mL}]_j}^2}\\
		\leq& \mu_0s\cdot \mu r\sigma\cdot \fronorm{\bDelta_{\mR}}\cdot \fronorm{\mX_{\mL}}.
	\end{align*}
	Using the same argument, one has
	\begin{align*}
			\lab\la\calG\calAT\calA\calGT(\bDelta_{\mL}(\mRQ)^\tranH),\mX_{\mL}\mR^\tranH\ra\rab 
			&\leq \mu_0s\cdot \mu r\sigma\cdot \fronorm{\bDelta_{\mL}}\cdot \fronorm{\mX_{\mL}}, \\
			\lab\la\calG\calAT\calA\calGT(\mL\bDelta_{\mR}^\tranH),\mL\mX_{\mL}^\tranH\ra\rab  
			&\leq  \mu_0s\cdot \mu r\sigma\cdot \fronorm{\bDelta_{\mR}}\cdot \fronorm{\mX_{\mL}},\\
			\lab\la\calG\calAT\calA\calGT(\bDelta_{\mL}(\mRQ)^\tranH),\mL\mX_{\mR}^\tranH\ra\rab  
			&\leq\mu_0s\cdot \mu r\sigma\cdot \fronorm{\bDelta_{\mL}}\cdot \fronorm{\mX_{\mR}}.
	\end{align*}

	Hence we can obtain
	\begin{align*}
		\lab\la\nabla f_1, \mX\ra\rab^2 
		\leq& \big(\mu_0\mu s r \sigma)^2\cdot(\fronorm{\bDelta_R}\fronorm{\mX_L} + \fronorm{\bDelta_L}\fronorm{\mX_L} + \fronorm{\bDelta_R}\fronorm{\mX_R} + \fronorm{\bDelta_L}\fronorm{\mX_R}\big)^2\\
		= &(\mu_0\mu s  r\sigma)^2\cdot (\fronorm{\bDelta_L} + \fronorm{\bDelta_R})^2\cdot(\fronorm{\mX_L} + \fronorm{\mX_R})^2\\
		\leq& 4(\mu_0\mu s r\sigma)^2\fronorm{\bDelta}^2,
	\end{align*}
	where the last line is due to $\fronorm{\mX}^2=\fronorm{\mX_L}^2 + \fronorm{\mX_R}^2 =1$. Thus we have 
	\begin{align*}
		\fronorm{\nabla f_1}^2  
		&=\sup_{\fronorm{\mX}=1} \lab \la\nabla f_1, \mX\ra\rab ^2 \\
		&\leq 4(\mu_0\mu s r\sigma)^2\fronorm{\bDelta}^2 .\numberthis\label{ineq: upper bound of fronorm of gls}
	\end{align*}
	\paragraph{Bounding $\fronorm{  \nabla f_2}^2$}
	For any $\mX\in \C^{(sn_1+n_2)\times r}$ such that $\fronorm{\mX}=1$, one has
	\begin{align*}
		|\la \nabla f_2, \mX \ra| 
		\leq &\left| \la (\calI - \calG\calGT)(\mL\mR^\tranH), \mX_{\mL}\mR^\tranH + \mL\mX_{\mR}^\tranH\ra \right|  \\
		=& \left| \la (\calI - \calG\calGT)(\mL\mR^\tranH - \mL^\natural{\mR^\natural}^\tranH), \mX_{\mL}\mR^\tranH + \mL\mX_{\mR}^\tranH\ra \right|\\
		\leq &\fronorm{ (\calI - \calG\calGT)(\mL\mR^\tranH - \mL^\natural{\mR^\natural}^\tranH) }\cdot \fronorm{  \mX_{\mL}\mR^\tranH + \mL\mX_{\mR}^\tranH} \\
		\leq&  \fronorm{  \mL\mR^\tranH - \mL^\natural{\mR^\natural}^\tranH  } \cdot  \fronorm{  \mX_{\mL}\mR^\tranH + \mL\mX_{\mR}^\tranH}.
	\end{align*}
	We bound both terms separately.
	\begin{itemize}
		\item Bounding $\fronorm{  \mL\mR^\tranH - \mL^\natural{\mR^\natural}^\tranH  }$. It can be bounded as follows:
		\begin{align*}
			\fronorm{  \mL\mR^\tranH - \mL^\natural{\mR^\natural}^\tranH  }&=\fronorm{\mL\bDelta_{\mR}^\tranH + \bDelta_{\mL}(\mRQ)^\tranH}\\
			&\leq \opnorm{\mL} \fronorm{\bDelta_{\mR}} + \opnorm{\mR^\natural}\fronorm{\bDelta_{\mL }} \\
			&\leq (1+\varepsilon )\sqrt{\sigma_1}\fronorm{\bDelta_{\mR}}  + \sqrt{\sigma_1} \fronorm{\bDelta_{\mL }}\\
			&\leq (1+\varepsilon)\sqrt{\sigma_1 } \cdot \sqrt{2} \fronorm{\bDelta},
		\end{align*}
		where we have used the fact $\opnorm{\mL} \leq \opnorm{\bDelta_{\mL }} + \opnorm{\mL^\natural} \leq \fronorm{\bDelta_{\mL}} + \sqrt{\sigma_1} \leq \fronorm{\bDelta} + \sqrt{\sigma_1} \leq (1+\varepsilon )\sqrt{\sigma_1}$.
		\item Bounding $\fronorm{  \mX_{\mL}\mR^\tranH + \mL\mX_{\mR}^\tranH}$. It can be bounded as follows:
		\begin{align*}
			\fronorm{  \mX_{\mL}\mR^\tranH + \mL\mX_{\mR}^\tranH}&\leq \opnorm{\mR}\fronorm{\mX_{\mL}} + \opnorm{\mL}\fronorm{\mX_{\mR}} \\
			&\leq (1+\varepsilon)\sqrt{\sigma_1} (\fronorm{\mX_{\mL} } + \fronorm{\mX_{\mR}})\\
			&\leq \sqrt{2}(1+\varepsilon)\sqrt{\sigma_1}.
		\end{align*}
	\end{itemize}
	Combining together and simple computation yields that 
	\begin{align}
		\label{ineq: upper bound of gradient of gh}
		\fronorm{\nabla f_2}^2 &=\sup_{\fronorm{\mX}=1} \lab \la\nabla f_3, \mX \ra \rab^2 \notag\\
		&\leq 4(1+\varepsilon)^4\sigma_1^2 \cdot \fronorm{\bDelta}^2.
	\end{align} 
	\paragraph{Bounding $\fronorm{\nabla f_3 }^2$} Recall that $$\mS=\begin{bmatrix}
		 \mI_{sn_1} &\\
		 &-\mI_{n_2}\\
	\end{bmatrix}.$$
	A straightforward computation yields that 
	\begin{align*}
		\fronorm{\nabla f_3  }^2
		=&\frac{1}{16} \left( \fronorm{\mL(\mL^\tranH\mL -\mR^\tranH\mR)}^2 + \fronorm{\mR(\mR^\tranH\mR \mR^\tranH\mR)}^2 \right) \\
		=&\frac{1}{16}\fronorm{\mS\mM\mM^\tranH\mS\mM}^2\\
		=& \frac{1}{16}\fronorm{\mS(\mM\mM^\tranH - \mM^\natural{\mM^\natural}^\tranH)\mS\mM + \mS\mM^\natural{\mM^\natural}^\tranH\mS\mM}^2\\
		\leq &\frac{1}{8}\fronorm{\mS(\mM\mM^\tranH - \mM^\natural{\mM^\natural}^\tranH)\mS\mM}^2  + \frac{1}{8}\fronorm{\mS\mM^\natural{\mM^\natural}^\tranH\mS\mM}^2\\
		 \leq& \frac{1}{8}\opnorm{\mM}^2\cdot\fronorm{\mM\mM^\tranH - \mM^\natural{\mM^\natural}^\tranH}^2  + \frac{1}{8}\opnorm{\mM^\natural}^2\cdot \fronorm{{\mM^\natural}^\tranH\mS(\mM^\natural+\bDelta)}^2\\
	    \stackrel{(a)}{=} &\frac{1}{8}\opnorm{\mM}^2\cdot\fronorm{\mM^\natural \bDelta^\tranH + \bDelta{\mM^\natural}^\tranH + \bDelta\bDelta^\tranH}^2 + \frac{1}{8}\opnorm{\mM^\natural}^2\cdot\fronorm{{\mM^\natural}^\tranH\mS\bDelta}^2\\
		\leq & \frac{3}{8}\opnorm{\mM}^2\cdot\left(2\opnorm{\mM^\natural}^2 \fronorm{\bDelta}^2+ \fronorm{\bDelta}^4 \right)+ \frac{1}{8}\opnorm{\mM^\natural}^2\cdot\fronorm{{\mM^\natural}^\tranH\mS\bDelta}^2,
	\end{align*}
	where step (a) follows from ${\mM^\natural}^\tranH\mS\mM^\natural = \bzero$.  Notice that $\opnorm{\mM^\natural} \leq \sqrt{2\sigma_1(\mZ^\natural)}$ and $\opnorm{\mM} \leq \fronorm{\mM-\mM^\natural\mQ} + \opnorm{\mM^\natural}\leq \sqrt{\frac{\varepsilon^2\sigma_r}{\mu_0 s}} + \sqrt{2\sigma_1}$. Thus we have
	\begin{align*}
		\fronorm{\nabla f_3  }^2
		\leq&~ \frac{3}{8}\lb\sqrt{\frac{\varepsilon^2\sigma_r}{\mu_0 s}} + \sqrt{2\sigma_1}\rb^2\cdot\lb 4\sigma_1 + \frac{\varepsilon^2\sigma_r}{\mu_0 s}\rb\fronorm{\bDelta}^2 + \frac{1}{4}\sigma_1\fronorm{{\mM^\natural}^\tranH\mS\bDelta}^2\\
		\leq&~ \frac{3}{8}\cdot 2(\varepsilon^2 + 2)\sigma_1\cdot (4+\varepsilon^2)\sigma_1\cdot \fronorm{\bDelta}^2+ \frac{1}{4}\sigma_1\fronorm{{\mM^\natural}^\tranH\mS\bDelta}^2\\
		\leq  &~\frac{19}{4} \sigma_1^2\cdot \fronorm{\bDelta}^2 + \frac{1}{4}\sigma_1\fronorm{{\mM^\natural}^\tranH\mS\bDelta}^2,\numberthis\label{ineq: upper bound of gradient of gb}
	\end{align*}
	 where the last line is due to $\varepsilon \leq \frac{1}{3}$.

Finally, plugging \eqref{ineq: upper bound of fronorm of gls}, \eqref{ineq: upper bound of gradient of gh} and \eqref{ineq: upper bound of gradient of gb} into \eqref{ineq: upper bound of gradient of fronorm of g}, we obtain that
\begin{align*}
	\fronorm{\nabla f(\mM)}^2
	 \leq &4\cdot 4(\mu_0\mu s r\sigma)^2\fronorm{\bDelta}^2 + 4\cdot  4(1+\varepsilon)^4\sigma_1^2 \cdot \fronorm{\bDelta}^2     + 2\left(\frac{19}{4} \sigma_1^2\cdot \fronorm{\bDelta}^2 + \frac{1}{4}\sigma_1\fronorm{{\mM^\natural}^\tranH\mS\bDelta}^2\right)\\
	\leq& \left( 64(\mu_0 \mu sr)^2 + 16(1+\varepsilon)^4 + \frac{19}{2}\right)\sigma_1^2 \fronorm{\bDelta}^2 + \frac{1}{2}\sigma_1\fronorm{{\mM^\natural}^\tranH\mS\bDelta}^2  \\
	\leq &125(\mu_0\mu s r \sigma_1)^2\fronorm{\bDelta}^2 + \frac{1}{2}\sigma_1\fronorm{{\mM^\natural}^\tranH\mS\bDelta}^2,\numberthis\label{ineq: upper bound of gradient of g}
\end{align*}
where the second line is due to $\sigma \leq \frac{(1+\varepsilon)\sigma_1}{1-\varepsilon}\leq 2\sigma_1$ and the last line follows from $\varepsilon\leq \frac{1}{3}< 1$. 

\subsection{Auxiliary Lemmas}
\label{sec: auxiliary}
\begin{lemma}[\cite{tropp2012user}{, Theorem 6.1}]
	\label{lem: matrix bernstein}
	Assume $\{\mX_i\}_{i=1}^n$ are independent random matrices of dimension $n_1\times n_2$ and obey $\E{\mX_i} = 0$ and $\opnorm{\mX_i}\leq B$. Define
	\begin{align*}
		\sigma^2:= \max\lcb\opnorm{\E{\sum_{i=1}^n \mX_i\mX_i^\tranH}}, \opnorm{\E{\sum_{i=1}^n \mX_i^\tranH\mX_i}}\rcb.
	\end{align*}
	Then the event
	\begin{align}
		\label{ineq: bernstein}
		\opnorm{\sum_{i=1}^n \mX_i} \leq c\lb\sqrt{\sigma^2\log(n_1+n_2)}+B\log(n_1+n_2)\rb
	\end{align}
	holds with probability at least $1-(n_1+n_2)^{-c_1}$, where $c,c_1>0$ are absolute constants.
\end{lemma}
\begin{lemma}[\cite{chen2020vectorized}{, Lemma III.13}]
	\label{lem: two norms}
	Suppose $\mZ^\natural$ is $\mu_1$-incoherent. Then one has
	\begin{align*}
		\sqrt{\sum_{i=0}^{n-1}\frac{\vecnorm{\calGT(\mZ)\ve_i}{2}^2}{\omega_i}} &\leq c_1\sqrt{\frac{\mu_1 r\log(sn)}{n}}\sigma_1(\mZ^\natural),\\
		\max_{0\leq i\leq n-1}\frac{\vecnorm{\calGT(\mZ)\ve_i}{2}}{\sqrt{\omega_i}} &\leq \frac{\mu_1 r}{n}\sigma_1(\mZ^\natural).
	\end{align*}
\end{lemma}

\begin{lemma}[\cite{chen2020vectorized}{, Corollary III.9}]
	\label{lem: pt opnorm}
	Under Assumption 1 and Assumption 2, let $T$ be the tangent space of $\mZ^\natural$, then the event
	\begin{align*}
		\opnorm{\calP_{T}\calG(\calAT\calA - \calI)\calGT\calP_{T}}\leq c \sqrt{\frac{\mu_0\mu_1 sr\log(sn)}{n}}	
	\end{align*}
	occurs with probability at least $1-(sn)^{-c}$ for some universal constant $c>0$.
\end{lemma}

\section{Conclusion}
\label{sec: conclusion}
In this paper, we propose a non-convex method called PGD--VHL for low rank vectorized Hankel matrix recovery problem in blind super-resolution of point sources. Our theoretical analysis shows that PGD--VHL converges to the target matrix linearly when the number of samples is larger than $\calO (s^2r^2 \log^2(sn))$.
The performance of PGD--VHL has also been demonstrated by our numerical simulations.
For future work, it is interesting to study the recovery performance of PGD--VHL in the presence of noise and the behavior of vanilla gradient descent method for blind super-resolution.

\section{Acknowledgment}
The authors would like to thank the anonymous reviewers and the Associate Editor for their useful comments which have helped to improve the quality of the present work. The authors also thank Ke Wei for helpful discussions.

\bibliographystyle{plain}
\bibliography{refs}

\end{document}